\documentclass[aps,pre,preprint,superscriptaddress,compress]{revtex4-2}

\usepackage{mathtools,booktabs}
\usepackage{bm}
\usepackage{autobreak}
\usepackage{amsmath}
\usepackage{amsthm}
\usepackage{siunitx}
\usepackage{gensymb}
\usepackage{subfigure}
\usepackage{color}
\usepackage{hyperref}
\usepackage{url}
\usepackage{CJKutf8}

\begin{document}
	\begin{CJK*}{UTF8}{gbsn}
		\preprint{APS}
		\title{A phase-field-based lattice Boltzmann method for two-phase flows with the interfacial mass/heat transfer}
		\author{Baihui Chen (陈百慧)}
		\affiliation{School of Mathematics and Statistics, Huazhong University of Science and Technology, Wuhan, 430074, China}
		\affiliation{Institute of Interdisciplinary Research for Mathematics and Applied Science, Huazhong University of Science and Technology, Wuhan 430074, China}
		\affiliation{Hubei Key Laboratory of Engineering Modeling and Scientific Computing, Huazhong University of Science and Technology, Wuhan, 430074, China}
		\author{Chengjie Zhan (湛承杰)}
		\affiliation{School of Mathematics and Statistics, Huazhong University of Science and Technology, Wuhan, 430074, China}
		\affiliation{Institute of Interdisciplinary Research for Mathematics and Applied Science, Huazhong University of Science and Technology, Wuhan 430074, China}
		\affiliation{Hubei Key Laboratory of Engineering Modeling and Scientific Computing, Huazhong University of Science and Technology, Wuhan, 430074, China}
		\author{Zhenhua Chai (柴振华)}
		\email[Corresponding author: ]{hustczh@hust.edu.cn}
		\affiliation{School of Mathematics and Statistics, Huazhong University of Science and Technology, Wuhan, 430074, China}
		\affiliation{Institute of Interdisciplinary Research for Mathematics and Applied Science, Huazhong University of Science and Technology, Wuhan 430074, China}
		\affiliation{Hubei Key Laboratory of Engineering Modeling and Scientific Computing, Huazhong University of Science and Technology, Wuhan, 430074, China}
		\author{Baochang Shi (施保昌)}
		\affiliation{School of Mathematics and Statistics, Huazhong University of Science and Technology, Wuhan, 430074, China}
		\affiliation{Institute of Interdisciplinary Research for Mathematics and Applied Science, Huazhong University of Science and Technology, Wuhan 430074, China}
		\affiliation{Hubei Key Laboratory of Engineering Modeling and Scientific Computing, Huazhong University of Science and Technology, Wuhan, 430074, China}
		
		\date{\today}
	\begin{abstract} 
		In this work, we develop a phase-field-based lattice Boltzmann (LB) method for a two-scalar model of the two-phase flows with interfacial mass/heat transfer. Through the Chapman–Enskog analysis, we show that the present LB method can correctly recover the governing equations for phase field, flow field and concentration/temperature field. In particular, to derive the two-scalar equations for the mass/heat transfer, we propose a new LB model with an auxiliary source distribution function to describe the extra flux terms, and the discretizations of some derivative terms can be avoided. The accuracy and efficiency of the present method is also tested through several benchmark problems, and the influence of mass/heat transfer on the fluid viscosity is further considered by introducing an exponential relation. The numerical results show that the present LB method is suitable for the two-phase flows with interfacial mass/heat transfer.
	\end{abstract}
	
	\maketitle
\end{CJK*}
		
\section{Introduction}
Two-phase flows with the interfacial mass/heat transfer are widely encountered in many scientific and engineering fields, including chemical reactors \cite{reactors1, reactors2}, boiling \cite{boiling1, boiling2}, PEM electrolysis cell \cite{fuel}, combustion \cite{combustion_two-phase_flow}, and carbon sequestration \cite{CO2multi}. These complex problems involve a range of physical processes that occur at different scales and include multiple fields, and it is difficult to obtain the analytical solutions. With the development of the computer techniques and scientific computing, the numerical simulation has become an essential tool for these problems \cite{review1, review2, review3}, and some different numerical methods have also been developed to capture the details at the interface of two-phase flows \cite{xu2022incompressible, devals2007two}. The key to investigate the interfacial mass/heat transfer lies in threefold: (a) capturing or tracking the fluid interface, (b) describing the convection and diffusion of mass/heat transfer in the bulk phase, and (c) determining the mass/heat transport at the interface.

Generally speaking, the methods for capturing or tracking the fluid interface can be divided into two main kinds, i.e., the sharp-interface and diffuse-interface methods. In the sharp-interface method, for instance, the volume of fluid (VOF) method \cite{vof, rider1998reconstructing} and the level set method \cite{levelset1, levelset2}, the interface thickness is assumed to be zero, which may be not true since the interface usually has a thin but non-zero thickness for the two-phase flows in reality. On the contrary, in the diffuse-interface method, the interface is assumed to have a finite but non-zero thickness, and all physical variables (e.g., density and velocity) and parameters (e.g., viscosity and diffusivity) change smoothly across the interface. As one of the diffuse-interface methods, the phase-field model has been widely used in the simulation of multiphase flows \cite{anderson1998diffuse, kim2012phase, santra2020phase, phase-field_review} since it does not need to track interface explicitly, and simultaneously, it can preserve mass conservation. 
Two-phase problems with the mass/heat transfer involve disparate lengths and time scales, and the scalar quantities often experience very large diffusivity ratios between different phases \cite{one-scalar1, one-scalar2}, which may lead to an unphysical leakage (artificial diffusion) at the interface and negative concentration \cite{jain2020modeling}. 
To overcome the artificial diffusion, a consistent treatment on the advection term of both discontinuities should be introduced, which can ensure the interface and the species concentration discontinuity surface are advected synchronously and preserve the positivity of scalars \cite{Davidson_Rudman_2002, alke2009vof, Bothe_Fleckenstein_2013}. 
However, most works are limited to the VOF method, and the studies on diffuse-interface method for scalar transport are relatively rare. Recently, Mirjalili et al. \cite{Mirjalili2022A} proposed a phase-field-based two-scalar model for the two-phase problems including the interfacial mass/heat transfer with arbitrary diffusivity ratios, and this model can prevent the unphysical leakage of mass/heat at the interface by preserving the thermodynamic equilibrium of the steady-state solution.

Over the past three decades, the lattice Boltzmann (LB) method has become a powerful computational tool in the study of complex fluid flows \cite{kruger2017lattice, Chen_1998, Aidun_2010}, such as the multiphase and multicomponent flows \cite{Abadi_2018, Zheng_2020, Yuan_2022, Chai_2019}, dendritic growth \cite{Sundongke_2019, Zhancj_2023}, phase transition \cite{Zhaoyong_2019, PhysRevLett.86.3578} and the fluid flows in porous media \cite{ChaiZH_2019, pan2004lattice}, due to its features of kinetic background, easy implementation of boundary conditions and fully parallel algorithm. Recently, the LB method has also been adopted to investigate the coupling problems of interfacial mass/heat transfer in two-phase flows \cite{worner2012numerical, mass_lb2, riaud2014lattice, zhao2015simulation}. To accurately describe the concentration or temperature scalar at the interface, Lu et al. \cite{Lu_2019} analyzed the relation between Henry's law and the weight coefficients in LB model for conjugate mass transfer, while the selection of parameters is restricted. Inspired by the continuum species transfer (CST) method \cite{HAROUN20102896}, Yang et al. \cite{Yang_2021} directly added the CST model as a source term into the convection-diffusion equation, but the mass flux continuity at the interface is established in the VOF framework. Later, Tan et al. \cite{Tan_2022} adopted a unified single-field concentration equation based on the phase-field method in Ref. \cite{Deising_2016}, and considered the gas contraction at the gas-liquid interface by introducing a source term. Recently, a diffuse-interface LB model for surfactant transport was proposed in Ref. \cite{Hu_2021_surfactant}, which has been further extended to solve a unified convection-diffusion equation with the Neumann boundary conditions in complex geometries \cite{Hu_2023_Neumann}. However, all above works were carried out based on one-scalar models, which still has a special limitation on the diffusivity ratio and may bring an unphysical leakage when the ratio of diffusivity is large.

In this work, we will propose a phase-field-based LB method  for the two-phase problems with interfacial mass/heat transfer based on the two-scalar model developed by Mirjalili et al. \cite{Mirjalili2022A}. In Section \ref{GoverEqs}, the governing equations for two-phase flows with interfacial mass/heat transfer are introduced, including the Allen-Cahn equation for phase field, the Navier-Stokes equations for flow field, the two-scalar equations as well as the degenerate one-scalar equation for mass/heat transfer. In Section \ref{LBMs}, the LB method for the governing equations are designed, and through the Chapman–Enskog analysis \cite{Chen_1998, Chapman-Enskog}, the two-scalar model can be correctly recovered from the present LB method. In Section \ref{Numerical}, the numerical validations and discussion are performed through several benchmark problems, and finally, some conclusions are summarized in Section \ref{Conclusion}.

\section{Governing equations for interfacial mass/heat transfer in two-phase flows}\label{GoverEqs}
In this section, we introduce the governing equations for the interfacial mass/heat transfer in two-phase flows. Specifically, the conservative Allen-Cahn equation \cite{Allen-Cahn1} is employed to describe the phase field, and the incompressible Navier-Stokes equations are utilized to describe the flow behavior of the complex system. Additionally, the concentration field or temperature field is characterized by using either the one-scalar or two-scalar model proposed by Mirjalili et al. \cite{Mirjalili2022A}.
\subsection{Allen-Cahn equation for phase field}
Based on the phase-field theory, the Allen-Cahn equation for the interface capturing can be expressed by \cite{Allen-Cahn1}
\begin{equation}\label{1.2}
	\frac{\partial \phi}{\partial t}+\nabla \cdot \left(\phi\mathbf{u}\right)=\nabla \cdot \left[ M \left( \nabla \phi - \frac{4\phi\left(1-\phi\right)}{W}\mathbf n \right) \right],
\end{equation}
where $\phi$ is the order parameter with $\phi=1$ in fluid 1 and $\phi=0$ in fluid 2, $\mathbf u$ is the fluid velocity, $M$ is the mobility coefficient, $W$ is the interface thickness, and $\mathbf{n}=\nabla\phi/|\nabla\phi|$ is the unit normal vector. Usually the initial distribution of the order parameter is approximated by the following hyperbolic tangent profile,
\begin{equation}\label{initial_phi}
	\phi(\mathbf x)= \frac{1}{2} + \frac{1}{2} \tanh\frac{2l\left(\mathbf x\right)}{W},
\end{equation}
where $l\left(\mathbf x\right)$ is a signed-distance function. 
\subsection{Incompressible Navier-Stokes equations for flow field}
To describe the fluid flows, we consider the following Navier-Stokes equations for incompressible Newtonian fluids, 
\begin{subequations}\label{1.3}
	\begin{equation}
		\nabla \cdot \mathbf{u}=0,
		\end{equation}
		\begin{equation}\label{ns_2}
		\frac{\partial\left(\rho \mathbf{u}\right)}{\partial t}+\nabla \cdot \left(\rho \mathbf{u} \mathbf{u}\right)=-\nabla p+\nabla \cdot \left[\mu \left(\nabla \mathbf{u}+ \nabla \mathbf{u}^T\right)\right]+\mathbf F_s +\mathbf G,
	\end{equation}
\end{subequations}
where $\rho$ is the fluid density, $p$ is the hydrodynamic pressure, $\mu$ is the dynamic viscosity, $\mathbf F_s=\mu_{\phi}\nabla \phi$ is the surface tension force, and $\mathbf G$ is a body force, such as gravity $\rho\mathbf{g}$. Here $\mu_{\phi}$ is the chemical potential defined by
\begin{equation}\label{nabla_2}
	\mu_{\phi}=4\beta \phi \left(\phi-1\right)\left(\phi-0.5\right)-k\nabla ^2 \phi,
\end{equation}
where $\beta$ and $k$ are physical parameters that depend on the interface thickness $W$ and the surface tension coefficient $\sigma$,
\begin{equation}
	\beta=\frac{12\sigma}{W}, k=\frac{3\sigma W}{2}.
\end{equation}
The density $\rho$ in Eq. (\ref{ns_2}) is usually a linear function of the order parameter $\phi$,
\begin{equation}
	\rho={\phi}\left(\rho_1-\rho_2\right)+\rho_2,
\end{equation}
where $\rho_1$ and $\rho_2$ are the densities of the pure fluid 1 and fluid 2, respectively. In addition, the dynamic viscosity $\mu$ is also a linear combination of viscosities $\mu_1$ and $\mu_2$ in fluids 1 and 2,
\begin{equation}\label{1.8}
	\mu=\phi \left(\mu_1-\mu_2\right)+\mu_2.
\end{equation}

\subsection{Two-scalar equations for mass/heat transfer}
The mass/heat evolution in a two-phase system includes the transport in the bulk of each phase and the mass/heat transfer at the two-phase interface. Usually, the mass transport in the bulk phases can be described by the following convection-diffusion equation,
\begin{equation}\label{0.1}
	\frac{\partial \widetilde{c_i}}{\partial t}+\nabla \cdot \left(\widetilde{c_i}\mathbf{u}\right)=\nabla \cdot \left(D_i\nabla \widetilde{ c_i}\right),
\end{equation}
where $i=1,2$ represents the phase index, $\widetilde{c_i}$ is the concentration and $D_i$ is the diffusivity in phase $i$. To satisfy the mass conservation in a closed system, the mass flux transfer at the interface should satisfy
\begin{equation}\label{0.2}
	D_1\nabla_n \widetilde{c_1}=D_2\nabla_n \widetilde{c_2}.
\end{equation}
Similarly, the governing equation for heat transport in the bulk phases is given by
\begin{equation}\label{0.11}
	\frac{\partial \widetilde{q_i}}{\partial t}+\nabla \cdot \left(\widetilde{q_i}\mathbf{u}\right)=\nabla \cdot \left(\frac{k_i}{\rho_i C_{p,i}}\nabla \widetilde{q_i}\right),
\end{equation}
where $k_i$ is the heat conductivity, $C_{p,i}$ is the specific heat capacity. The heat balance at the interface is given by
\begin{equation}\label{0.21}
	\frac{k_1}{\rho_1 C_{p,1}} \nabla_n \widetilde{q_1}=\frac{k_2}{\rho_2 C_{p,2}}\nabla_n \widetilde{q_2}.
\end{equation}
Due to the chemical/thermal equilibrium at the interface, one can obtain the following relations,
\begin{equation}\label{0.31}
	\frac{\widetilde{c_1}}{\widetilde{c_2}}=H,\, \frac{\widetilde{q_1}}{\widetilde{q_2}}=\frac{\rho_1 C_{p,1}}{\rho_2 C_{p,2}},
\end{equation}
where $H$ is the Henry coefficient in Henry's law.
\par Take two-dimensional case as an example, we can obtain the following equation from Eq. (\ref{0.1}) through introducing $\overline{\widetilde{c_i}}$ being the average of concentration in $y$ direction,   
\begin{equation}\label{0.3}
	\frac{\partial \overline {\widetilde {c_i}} }{\partial t}+\frac{\partial \left( \overline{\widetilde{c_i}}\mathbf{u}\right)}{\partial x}=\frac{\partial}{\partial x}\left( D_i \frac{\partial \overline{\widetilde{c_i}}}{\partial x}\right).
\end{equation}
The $y$-averaged mean concentration values in each phase are then denoted by $\overline{\widetilde{c_1}}(x)={c_1(x)}/{\phi(x)}$ and $\overline{\widetilde{c_2}}(x)={c_2(x)}/\left({1-\phi(x)}\right)$. Based on this assumption, Eq. (\ref{0.3}) can be further rewritten as
\begin{subequations}\label{simplify-eq}
	\begin{equation}\label{0.6}
		\frac{\partial c_1}{\partial t}+\frac{\partial \left(c_1\mathbf{u}\right)}{\partial x}= \phi\frac{\partial}{\partial x}\left(D_1\frac{\partial \overline{\widetilde{c_1}}}{\partial x}\right), 
	\end{equation}
	\begin{equation}
		\frac{\partial c_2}{\partial t}+\frac{\partial \left(c_2\mathbf{u}\right)}{\partial x}= \left(1-\phi\right)\frac{\partial}{\partial x}\left(D_2\frac{\partial \overline{\widetilde{c_2}}}{\partial x}\right),
	\end{equation}
\end{subequations}
where the relation $\partial_t\phi+\mathbf{u} \cdot \nabla \phi=0$ is used. 
If the scalar only exists in phase 1 and does not diffuse into phase 2, i.e., $D_2=0$, we can derive $D_1 \partial_x\overline{\widetilde{c_1}} \partial_x \phi=0$ according to Eq. (\ref{0.2}), and similarly, we have $D_2 \partial_x\overline{\widetilde{c_2}} \partial_x \phi=0$ in phase 2. Therefore, Eq. (\ref{simplify-eq}) can be rewritten as
\begin{subequations}\label{0.7}
	\begin{equation}
		\frac{\partial c_1}{\partial t}+\frac{\partial \left(c_1\mathbf{u}\right)}{\partial x}= \frac{\partial}{\partial x}\left(D_1\phi\frac{\partial \overline{\widetilde{c_1}}}{\partial x}\right)=\frac{\partial}{\partial x} \left[D_1\left(\frac{\partial c_1}{\partial x}-\frac{4\left(1-\phi\right)c_1}{W}\mathbf{n}\right) \right], 
	\end{equation}
	\begin{equation}
		\frac{\partial c_2}{\partial t}+\frac{\partial \left(c_2\mathbf{u}\right)}{\partial x}= \frac{\partial}{\partial x}\left(D_2(1-\phi)\frac{\partial \overline{\widetilde{c_2}}}{\partial x}\right)=\frac{\partial}{\partial x} \left[D_2\left(\frac{\partial c_2}{\partial x}+\frac{4\phi c_2}{W}\mathbf{n}\right) \right],
	\end{equation}
\end{subequations}
where the following approximation for the gradient of order parameter is adopted,
\begin{equation}\label{0.8}
	\nabla \phi = \frac{4\phi\left(1-\phi\right)}{W}{\mathbf{n}}.
\end{equation}
To further describe the interfacial mass/heat transfer at fluid interface, the interface flux terms should be introduced into Eq. (\ref{0.7}), and the two-scalar equations can be given by \cite{Mirjalili2022A}
\begin{subequations}\label{1.1}
	\begin{equation}\label{two-scalar_1}
		\frac{\partial{c_1}}{\partial t}+\nabla \cdot\left(c_1\mathbf{u}\right)=\nabla \cdot \left[D_1\left(\nabla c_1-\frac{4(1-\phi)c_1}{W}\mathbf{n}\right) \right]+AD_m\left[K_{eq}c_2\phi-c_1\left(1-\phi\right)\right]-D_m\nabla\phi\cdot\nabla (c_1+K_{eq}c_2), 
	\end{equation}
	\begin{equation}
		\frac{\partial{c_2}}{\partial t}+\nabla \cdot\left(c_2\mathbf{u}\right)=\nabla \cdot \left[D_2\left(\nabla c_2+\frac{4\phi c_2}{W}\mathbf{n}\right) \right]+AD_m\left[c_1\left(1-\phi\right)-K_{eq}c_2\phi\right]+D_m\nabla\phi\cdot\nabla (c_1+K_{eq}c_2),
	\end{equation}
\end{subequations}
where $A$ is a free parameter representing an inverse time scale to thermodynamic equilibrium, $D_m$ is the mixed diffusivity and can be expressed as
\begin{equation}
	D_m=\frac{D_1D_2}{K_{eq}D_1(1-\phi)+D_2\phi}.
\end{equation}
Here $K_{eq}$ is used to represent an arbitrary jump in the equilibrium scalar concentration at the interface, $K_{eq}=H$ is for mass transfer while $K_{eq}={\rho_1 C_{p,1}}/{\rho_2 C_{p,2}}$ for heat transfer.
\subsection{One-scalar model for mass/heat transfer}
If we introduce $c=c_1+c_2$ to be the total species concentration content per total volume, then from Eq. (\ref{0.7}) and the thermodynamic equilibrium at the interface, i.e., $\overline{\widetilde{c_1}}=K_{eq}\overline{\widetilde{c_2}}$, the following consistent one-scalar model can be obtained, 
\begin{equation}\label{1.5}
	\frac{\partial{c}}{\partial t}+\frac{\partial(c\mathbf{u})}{\partial x}=\frac{\partial}{\partial x} \left[\left(D_1\phi K_{eq}+ D_2\left(1-\phi\right)\right)\frac{\partial \overline{\widetilde{c_2}}}{\partial x} \right].
\end{equation}
By using the relation $c=\overline{\widetilde{c_1}}\phi + \overline{\widetilde{c_2}}\left(1-\phi\right)$ and extending the above one-dimensional equation, a consistent one-scalar model for total species concentration content per total volume can be derived, 
\begin{equation}\label{31}
	\frac{\partial c}{\partial t}+\nabla \cdot (c\mathbf{u})= \nabla \cdot \left[D_{eff} \nabla \left(\frac{c}{K_{eff}} \right)\right],
\end{equation}
where the effective diffusivity is given by
\begin{equation}
	D_{eff}=D_1K_{eq}\phi+D_2(1-\phi),
\end{equation}
and the equilibrium concentration ratio across the interface is
\begin{equation}
	K_{eff}=K_{eq}\phi+(1-\phi).
\end{equation}

\section{Lattice Boltzmann method for the two-phase flows with interfacial mass/heat transfer}\label{LBMs}
In this section, we introduce several LB models for the governing equations introduced in above section. Usually, the LB method can be categorized into single-relaxation-time (SRT) \cite{1992LatticeBGK}, two-relaxation-time \cite{ginzburg2005equilibrium, ginzburg2008two}, and multiple-relaxation-time \cite{d1992generalized,Chapman-Enskog} models based on the different forms of collision term employed. Considering the computational efficiency, we adopt the SRT-LB model in this work.

\subsection{Lattice Boltzmann model for phase field}
Following the previous work \cite{liang_phase-field-based_2018}, the LB evolution equation for the phase-field model (\ref{1.2}) can be written as
\begin{equation}
	f_i\left(\mathbf{x}+\mathbf{c}_i \delta t,t+\delta t\right)=f_i\left(\mathbf{x},t\right)-\frac{1}{\tau_f}\left[f_i\left(\mathbf{x},t\right)-f_i^{eq}\left(\mathbf{x},t\right)\right]+\delta t\left(1-\frac{1}{2\tau_f}\right)F_i\left(\mathbf{x},t\right),
\end{equation}
where the $f_i(\mathbf{x},t)$ represents the particle distribution function at position $\mathbf{x}$ and $t$ along the discrete velocity direction $i$ ($i=0, 1, 2, \cdots, Q$ with $Q$ being the number of the discrete velocity), $f_i^{eq}(\mathbf x,t)$ is the corresponding equilibrium distribution function, $\tau_f$ represents the relaxation factor related to the mobility, $\mathbf c_i$ is the discrete velocity and $\delta t$ is the time step. For the Allen-Cahn equation (\ref{1.2}), the equilibrium distribution function $f_i^{eq}(\mathbf{x},t)$ can be defined as
\begin{equation}
	f_i^{eq}=\omega_i \phi \left( 1+\frac{\mathbf{c}_i \cdot \mathbf{u}}{c_s^2} \right),
\end{equation}
where $\omega_i$ is the weight coefficient, $c_s$ represents the lattice sound speed. In order to exactly recover the Allen-Cahn equation, the source distribution function $F_i(\mathbf{x},t)$ is given by
\begin{equation}
	F_i=\omega_i \left[\frac{\mathbf{c}_i \cdot \partial_t(\phi\mathbf{u})}{c_s^2}+\mathbf{c}_i \cdot \frac{4\phi(1-\phi)}{W}\mathbf n \right],
\end{equation}
and the order parameter $\phi$ can be calculated by
\begin{equation}
	\begin{aligned}
		\phi=\sum_i f_i.
	\end{aligned}
\end{equation}
In addition, through some asymptotic analysis methods, the mobility can be determined by
\begin{equation}
	M = {\delta t\left(\tau_f-\frac{1}{2}\right)c_s^2}.
\end{equation}

\subsection{Lattice Boltzmann model for flow field}
For the Naiver-Stokes equations for flow field, the LB evolution equation can be expressed as  \cite{liang_phase-field-based_2018}
\begin{equation}
	g_i\left(\mathbf x+\mathbf{c}_i\delta t,t+\delta t\right)=g_i\left(\mathbf x,t\right)-\frac{1}{\tau_g}\left[g_i\left(\mathbf x,t\right)-g_i^{eq}\left(\mathbf x,t\right)\right]+\delta t\left(1-\frac{1}{2\tau_g}\right)G_i\left(\mathbf x,t\right),
\end{equation}
where $g_i(\mathbf x,t)$ is the distribution function of fluid field, $\tau_g$ represents the dimensionless relaxation time relevant to the viscosity. $G_i(\mathbf x,t)$ is a distribution function related to the surface tension force and body force, $\tau_g$ represents the dimensionless relaxation time relevant to the viscosity.
$g_i^{eq}(\mathbf{x},t)$ represents the equilibrium distribution function, and can be designed as
\begin{equation}
	g_i^{eq}=\left\{
	\begin{aligned}
		&\frac{p}{c_s^2}\left(\omega_i-1\right)+\rho s_i(\mathbf{u}), & i=0, \\
		&\frac{p}{c_s^2}\omega_i+\rho s_i(\mathbf{u}), & i\neq0,
	\end{aligned}
	\right.
\end{equation}
where $s_i(\mathbf{u})$ is given by
\begin{equation}
	s_i(\mathbf{u})=\omega_i\left[\frac{\mathbf{c}_i\cdot \mathbf{u}}{c_s^2}+ \frac{{\left(\mathbf{c}_i\cdot \mathbf{u}\right)}^2}{2 c_s^4}-\frac{\mathbf{u}\cdot\mathbf{u}}{2 c_s^2} \right].
\end{equation}
In order to recover the incompressible Navier-Stokes equations, the force distribution function $G_i(x,t)$ can be defined as
\begin{equation}
	G_i=\omega_i\left[\frac{\mathbf{c}_i \cdot \left(\mu_{\phi}\nabla\phi+\mathbf G\right)}{c_s^2}+\frac{\left(\rho_1-\rho_2\right)\mathbf{u}\nabla \phi : \mathbf{c}_i\mathbf{c}_i}{c_s^2}\right].
\end{equation}
With the zeroth-order and first-order moments of the distribution function $g_i$, the pressure $p$ and macroscopic velocity $\mathbf{u}$ can be evaluated by
\begin{subequations}
	\begin{equation}
		\mathbf{u}=\frac{1}{\rho} \left[\sum_i\mathbf{c}_ig_i+0.5\delta t\left(\mu_{\phi}\nabla\phi+\mathbf G\right)\right],
		\end{equation}
		\begin{equation}	
		p=\frac{c_s^2}{1-\omega_0}\left[\sum_{i\neq0}g_i+\frac{\delta t}{2}(\rho_1-\rho_2)\mathbf{u}\cdot\nabla \phi +\rho s_0 (\mathbf{u})\right].
	\end{equation}
\end{subequations}
Finally, based on the Chapman-Enskog analysis, the fluid dynamic viscosity is given by
\begin{equation}
	\mu=\rho{\delta t\left(\tau_g-\frac{1}{2}\right)c_s^2}.
\end{equation}

\subsection{Lattice Boltzmann model for concentration/temperature field with two-scale equations}
To accurately consider the third terms in the right-hand side of Eq. (\ref{1.1}) at the order of the derivative operator, we will introduce the corresponding auxiliary source distribution function in the present LBM. The LB evolution equations for the two-scalar mass/heat model [Eq. (\ref{1.1})] read
\begin{subequations}
	\begin{equation}\label{LB_two_1}
		h_{1,i}\left(\mathbf x+\mathbf c_i\delta t,t+\delta t\right)=h_{1,i}\left(\mathbf x,t\right)-\frac{1}{\tau_{h_{1}}}\left[h_{1,i}\left(\mathbf x,t\right)-h_{1,i}^{eq}\left(\mathbf x,t\right)\right]+\delta t\left(1-\frac{1}{2\tau_{h_{1}}}\right)H_{1,i}\left(\mathbf x,t\right)+\delta t {\widetilde H_{1,i}}\left(\mathbf x,t\right),
	\end{equation}
	\begin{equation}
		h_{2,i}\left(\mathbf x+\mathbf c_i\delta t,t+\delta t\right)=h_{2,i}\left(\mathbf x,t\right)-\frac{1}{\tau_{h_{2}}}\left[h_{2,i}\left(\mathbf x,t\right)-h_{2,i}^{eq}\left(\mathbf x,t\right)\right]+\delta t\left(1-\frac{1}{2\tau_{h_{2}}}\right)H_{2,i}\left(\mathbf x,t\right)+\delta t {\widetilde H_{2,i}}\left(\mathbf x,t\right),
	\end{equation}
\end{subequations}
where $h_{1,i}(\mathbf{x},t)$ and $h_{2,i}(\mathbf{x},t)$ are the distribution functions of concentrations $c_1$ and $c_2$, $h_{1,i}^{eq}(\mathbf x,t)$ and $h_{2,i}^{eq}(\mathbf x,t)$ are the corresponding equilibrium distribution functions, $H_{1,i}(\mathbf x,t)$, $H_{2,i}(\mathbf x,t)$, $\widetilde H_{1,i}(\mathbf x,t)$, $\widetilde H_{2,i}(\mathbf x,t)$ are the distribution functions of the source terms in Eq. (\ref{1.1}), $\tau_{h_{1}}$ and $\tau_{h_{2}}$ represent the relaxation times related to the diffusivities. The equilibrium distribution function $h_{1,i}^{eq}(\mathbf{x},t)$ and $h_{2,i}^{eq}(\mathbf{x},t)$ can be presented as
\begin{subequations}
	\begin{equation}\label{h_equilibrium}
		h_{1,i}^{eq}=\omega_i c_1\left( 1+\frac{\mathbf{c}_i \cdot \mathbf{u}}{c_s^2} \right),
	\end{equation}
	\begin{equation}
	h_{2,i}^{eq}=\omega_i c_2\left( 1+\frac{\mathbf{c}_i \cdot \mathbf{u}}{c_s^2} \right).
	\end{equation}
\end{subequations}
Usually, when the LB model is applied for convection-diffusion (\ref{1.1}), the extra flux terms need to be discretized by some finite-difference schemes. In contrast, to overcome this problem, we introduce the following distribution functions $H_{1,i}(\mathbf x,t)$ and $H_{2,i}(\mathbf x,t)$ to deal with the source terms, and the distribution functions $\widetilde H_{1,i}(\mathbf x,t)$ and $\widetilde H_{2,i}(\mathbf x,t)$ to handle the interface flux terms in Eq. (\ref{1.1}), which are given by

\begin{subequations}\label{macro}
	\begin{equation}
		H_{1,i}=\omega_i\left(\frac{\mathbf{c}_i \cdot \partial_t\left(c_1\mathbf{u}\right)}{c_s^2}+\mathbf{c}_i \cdot\frac{4\left(1-\phi \right)c_1}{W}\mathbf{n}+ AD_m \left[K_{eq}c_2\phi-c_1\left(1-\phi\right)\right]\right),
	\end{equation}
	\begin{equation}
		H_{2,i}=\omega_i\left(\frac{\mathbf{c}_i \cdot \partial_t\left(c_2\mathbf{u} \right)}{c_s^2}-\mathbf{c}_i \cdot\frac{4\phi c_2}{W}\mathbf{n}+AD_m\left[c_1\left(1-\phi\right)-K_{eq}c_2\phi\right]\right),
	\end{equation}
	\begin{equation}\label{derivate_1}
		\widetilde H_{1,i}=-\omega_i D_m\nabla\phi\cdot\nabla \left(c_1+K_{eq}c_2 \right),
	\end{equation}
	\begin{equation}\label{derivate_2}
		\widetilde H_{2,i}=\omega_i D_m\nabla\phi\cdot\nabla \left(c_1+K_{eq}c_2 \right).
	\end{equation}
\end{subequations}

Through the Chapman-Enskog analysis shown in \ref{CE_analysis}, the macroscopic concentrations $c_1$ and $c_2$ can be calculated by the following equations,
\begin{subequations}
	\begin{equation}
		c_1=\sum_i h_{1,i}+\frac{\delta t}{2}AD_m \left[K_{eq}c_2\phi-c_1(1-\phi)\right], 
	\end{equation}
	\begin{equation}
		c_2=\sum_i h_{2,i}+\frac{\delta t}{2}AD_m\left[c_1(1-\phi)-K_{eq}c_2\phi\right].
	\end{equation}
\end{subequations}
The following relations between the diffusivities and dimensionless relaxation times can be derived,
\begin{subequations}
	\begin{equation}
		D_1={\delta t\left(\tau_{h_{1}}-\frac{1}{2}\right)c_s^2},
	\end{equation}
	\begin{equation}
		D_2={\delta t\left(\tau_{h_{2}}-\frac{1}{2}\right)c_s^2}.
	\end{equation}
\end{subequations}

\par Furthermore, it should be noted that the derivative terms in Eqs. (\ref{derivate_1}) and (\ref{derivate_2}) should be discretized with suitable difference schemes. By summing Eq. (\ref{23}) over $i$, we can derive a local computing scheme of $\nabla c_1$ and $\nabla c_2$ in the LB framework,
\begin{subequations}\label{39}
	\begin{equation}
		\nabla c_1= \frac{1}{\tau_{h_1}\delta t c_s^2}\left[-\sum \mathbf{c}_ih_{1,i}+c_1 \mathbf{u}-0.5\delta t\partial_{t}\left(c_1\mathbf{u}\right)+D_1\frac{4\left(1-\phi\right)c_1}{W}\frac{\nabla \phi}{|\nabla \phi|}\right],
	\end{equation}
	\begin{equation}
		\nabla c_2= \frac{1}{\tau_{h_2}\delta t c_s^2}\left[-\sum \mathbf{c}_ih_{2,i}+c_2 \mathbf{u}-0.5\delta t\partial_{t}\left(c_2\mathbf{u}\right)-D_2\frac{4\phi c_2}{W}\frac{\nabla \phi}{|\nabla \phi|}\right],
	\end{equation}
\end{subequations}
where the gradient of order parameter and its gradient norm can be also calculated by a local scheme \cite{wang2016comparative}, 
\begin{subequations}
	\begin{equation}
		|\nabla \phi|= \frac{-|C|-B}{D},
	\end{equation}
	\begin{equation}
		\nabla \phi= \frac{C}{D+B/|\nabla \phi|},
	\end{equation}
\end{subequations}
where $D=-\tau_f\delta t c_s^2$, $B=4M\delta t\phi \left(1-\phi\right)/W$, and $C=\sum \mathbf{c}_if_{i}-\phi\mathbf{u}+0.5\delta t\partial_{t}\left(\phi\mathbf{u}\right)$.
As an alternative, one can also use the following second-order isotropic central scheme for the gradient operator,
\begin{equation}
	\nabla \chi\left( \mathbf{x}\right)=\sum_{i\neq0}\frac{\omega_i\mathbf{c}_i\chi\left(\mathbf{x}+\mathbf{c}_i\delta t\right)}{c_s^2\delta t},
\end{equation}
where $\chi$ represents the order parameter $\phi$ and the concentrations $c_1$, $c_2$. In addition, the Laplace operator in Eq. (\ref{nabla_2}) is usually computed by
\begin{equation}
	\nabla^2 \phi\left( \mathbf{x}\right)=\sum_{i\neq0}\frac{2\omega_i\left[ \phi\left(\mathbf{x}+\mathbf{c}_i\delta t\right)-\phi\left(\mathbf{x}\right) \right]}{c_s^2\delta t^2}.
\end{equation}

\subsection{Lattice Boltzmann model for concentration/temperature field with one-scale equation}
The LB evolution equation for the one-scalar mass/heat model [Eq. (\ref{31})] can be expressed as
\begin{equation}
	h_{3,i}\left(\mathbf x+\mathbf c_i \delta t,t+\delta t\right)=h_{3,i}\left(\mathbf x,t\right)-\frac{1}{\tau_{h_{3}}}\left[h_{3,i}\left(\mathbf x,t\right)-h_{3,i}^{eq}\left(\mathbf x,t\right)\right]+\delta t\left(1-\frac{1}{2\tau_{h_{3}}}\right)H_{3,i}\left(\mathbf x,t\right),
\end{equation}
where the $h_{3,i}(\mathbf{x},t)$ is the concentration distribution function of total concentration $c$, $\tau_{h_{3}}$ represents the relaxation time, $h_{3,i}^{eq}(\mathbf x,t)$ is the equilibrium distribution function, which can be designed as
\begin{equation}
	h_{3,i}^{eq}=\left\{
	\begin{aligned}
		&c+\left(\omega_i-1\right){c}/{K_{eff}}, & i=0, \\
		&\omega_i {c}/{K_{eff}}+\omega_i {\mathbf c_i \cdot c\mathbf u}/{c_s^2}, & i\neq0,
	\end{aligned}
	\right.
\end{equation}
and $H_{3,i}(\mathbf x,t)$ is the auxiliary source distribution function, 
\begin{equation}
	H_{3,i}=\omega_i \frac{\mathbf {c}_i \cdot \partial_t(c\mathbf{u})}{c_s^2}.
\end{equation}
The macroscopic scalar concentration $c$ can be computed by
\begin{equation}
	c=\sum_i h_{3,i},
\end{equation}
and the effective diffusivity can be determined by the following relation
\begin{equation}
	D_{eff}=\delta t \left(\tau_{h_{3}}-\frac{1}{2}\right) c_s^2.
\end{equation}

\section{Numerical validations and discussion}\label{Numerical}	 
In this section, we first validate the accuracy of the proposed LB method, and the consistency between the two-scalar model and the one-scalar model by some benchmark problems. Then the present method is extended to study the more practical problems where the viscosity is related to concentration. In LB method, the D1Q3, D2Q9, and D3Q15 lattice structures are applied for the one-dimensional, two-dimensional, and three-dimensional problems \cite{1992LatticeBGK}.
In the following simulations, the half-way anti-bounce-back boundary scheme \cite{zhangT_bounce_back} is adopted for the Dirichlet boundary condition, the half-way bounce-back scheme \cite{ladd_1994_1} is applied for no-flux scalar and no-slip velocity boundary conditions, and the non-equilibrium extrapolation scheme \cite{zhao2002non} is adopted for pressure boundary condition. Due to the parallelism of LB method, the GPU-CUDA tool is adopted to improve the computational efficiency, and some parameters are fixed as $M=0.01$, $A=1000$, $W=4\Delta x$ without otherwise stated.

\subsection{One-dimensional tests}
In this part, we consider a one-dimensional static droplet in the domain $[-1,1]$ with different initial distributions and boundary conditions of the concentration or temperature. The grid size is set to be $N=200$, which leads to $\Delta x=0.01$. 
\subsubsection{Pseudo-single-phase case}
We first consider a pseudo-single-phase problem with the matched diffusivities on the two sides of the fluid interface, i.e., $D_1=D_2$, thus, the interface has no effect on the transport of the concentration throughout the whole domain. In this study, the physical parameters $D_1=D_2=1$ and $K_{eq}=1$, the scalar concentrations are initialized as $c_1(x, t=0)=5e^{-4{x}^2}\phi$ and $c_2(x, t=0)=5e^{-4{x}^2}\left(1-\phi\right)$, and the periodic boundary condition is imposed on the left and right boundaries. In addition, the signed-distance function is give by $l\left(\mathbf x\right)=-\left(x-0.5\right)\left(x+0.5\right)$. 

\begin{figure}
	\centering
	\includegraphics[width=4.0in]{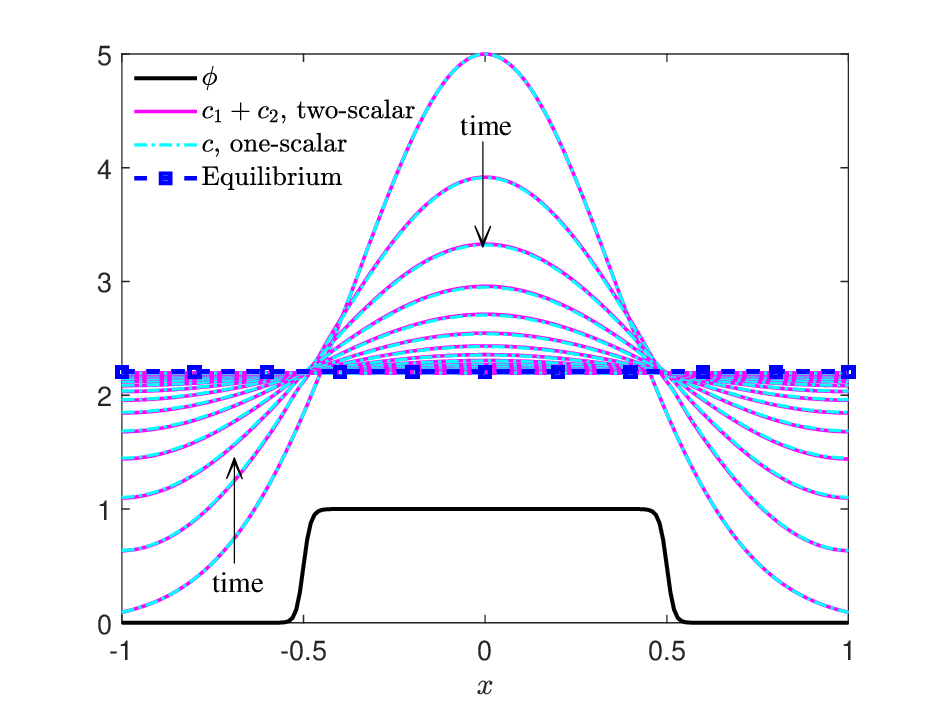}
	\caption{The distributions of two-scalar and one-scalar models at the same instants for pseudo-single-phase case, and the time intervals between adjacent concentration profiles are identical.}
	\label{fig-evolution}
\end{figure}

Figure \ref{fig-evolution} illustrates the profiles of the order parameter, the concentrations of one-scalar and two-scalar models, as well as the solution at equilibrium state.
From this figure, one can find that the numerical results of one-scalar and two-scalar models are consistent with each other, and they are in good agreement with the analytical solution at the equilibrium state, which indicates that the present LB method is accurate for both one-scalar and two-scalar models.

\subsubsection{Flat equilibrium case}\label{flat_section}
When the concentrations are initialized by $c_1\left(x,t=0\right)=2\phi$ and $c_2\left(x,t=0\right)=0$, $l\left(\mathbf{x}\right)=-\left(x-0.5\right)\left(x+0.5\right)$ and the periodic boundary condition is used, the equilibrium profiles of $c_1$ and $c_2$ are the flat inside and outside of the drop with a jump controlled by parameter $K_{eq}$, and $c_1+K_{eq}c_2$ should also keep a flat profile in the whole domain. 

\begin{figure}
	\centering
	\subfigure[]
	{
		\begin{minipage}{0.45\linewidth}		
			\centering		
			\includegraphics[width=3.0in]{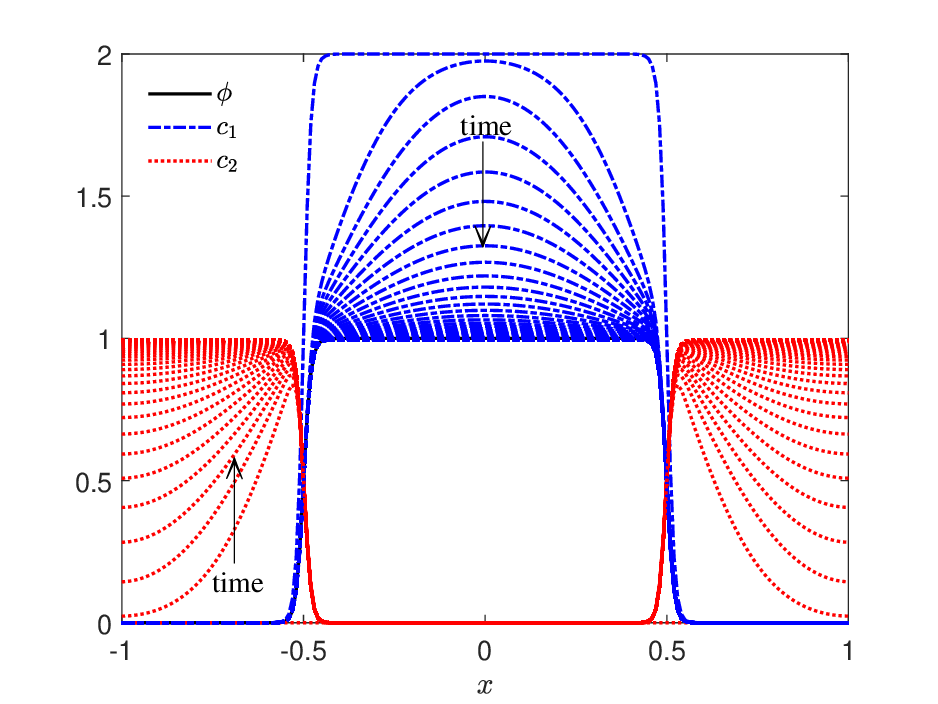}
		\end{minipage}
	}
	\subfigure[]
	{
		\begin{minipage}{0.45\linewidth}
			\centering
			\includegraphics[width=3.0in]{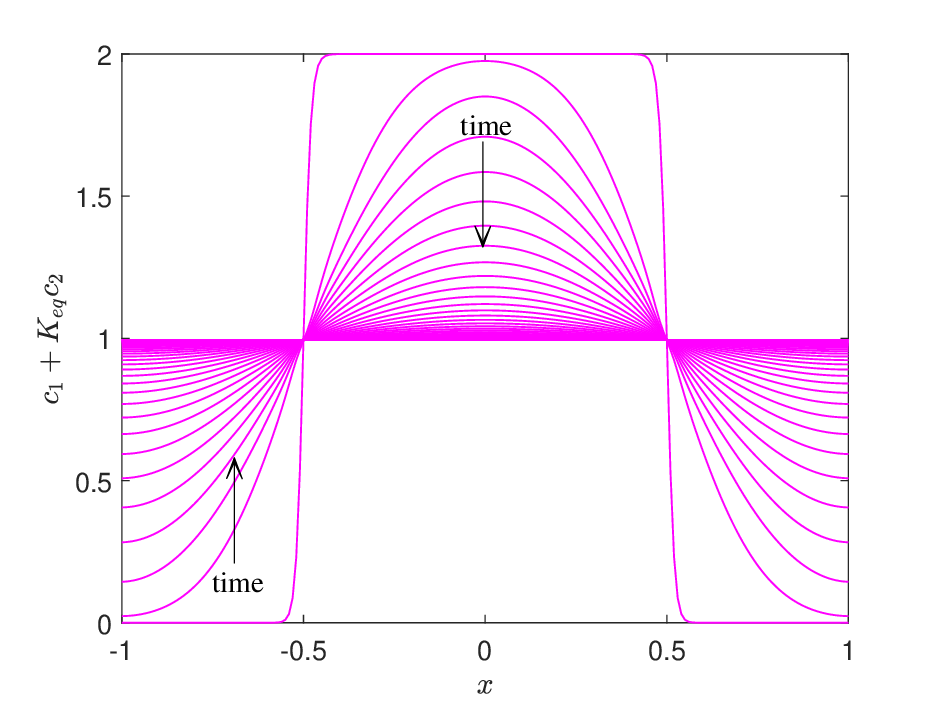}
		\end{minipage}
	}
	\subfigure[]
	{
		\begin{minipage}{0.45\linewidth}		
			\centering		
			\includegraphics[width=3.0in]{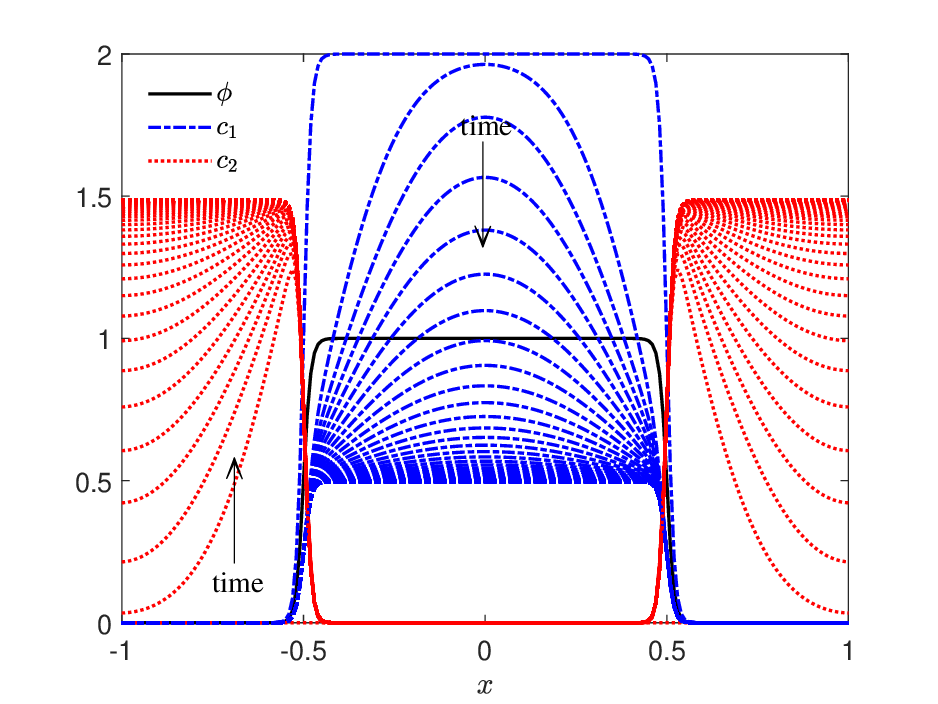}
		\end{minipage}
	}
	\subfigure[]
	{
		\begin{minipage}{0.45\linewidth}
			\centering
			\includegraphics[width=3.0in]{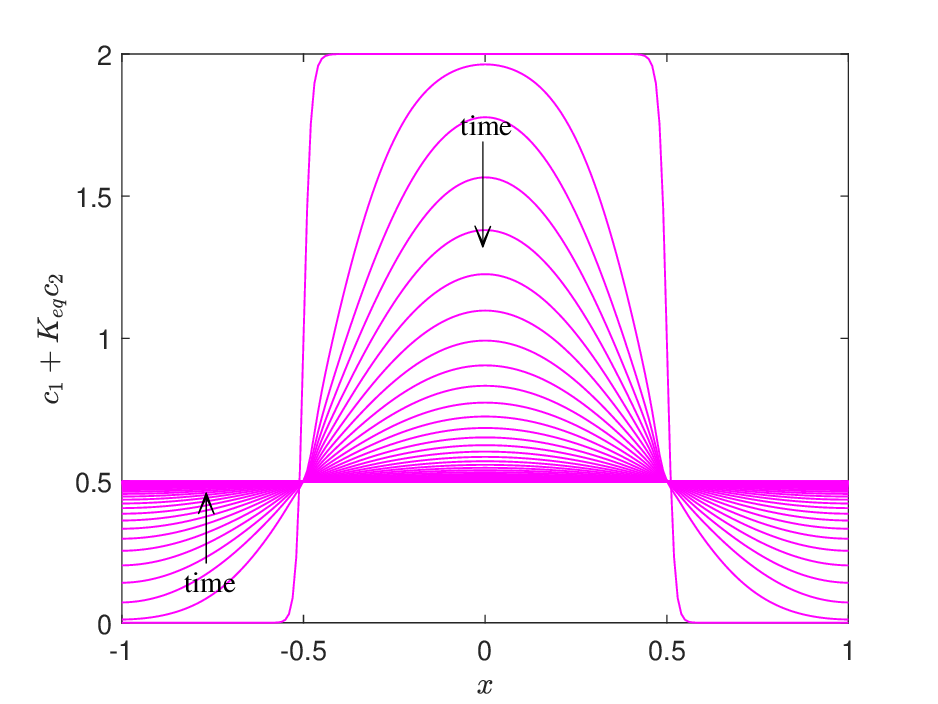}
		\end{minipage}
	}
	\subfigure[]
	{
		\begin{minipage}{0.45\linewidth}
			\centering
			\includegraphics[width=3.0in]{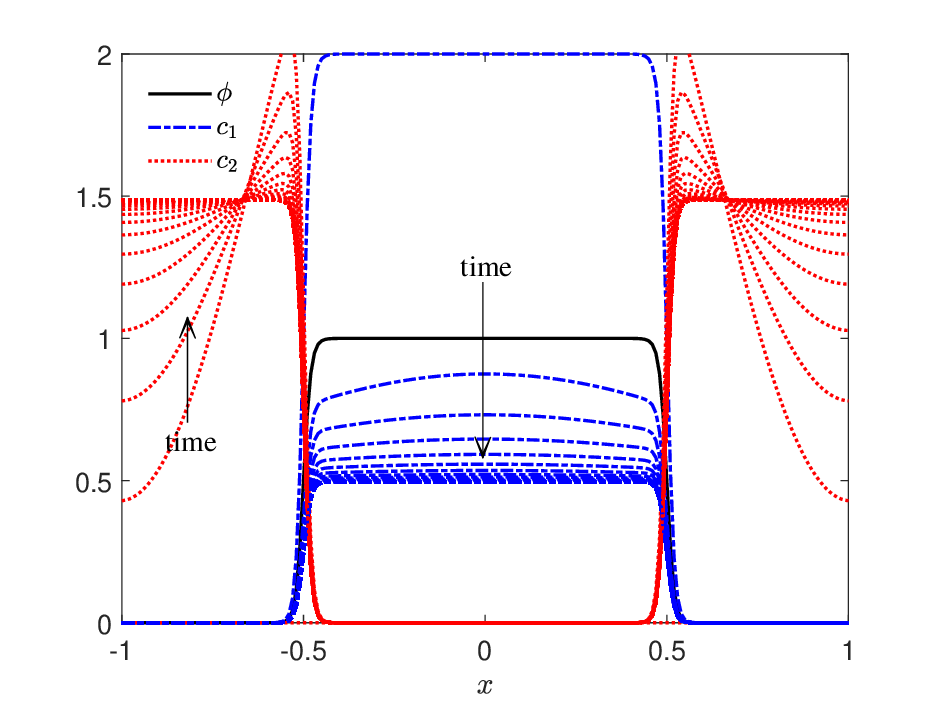}
		\end{minipage}
	}
	\subfigure[]
	{
		\begin{minipage}{0.45\linewidth}
			\centering
			\includegraphics[width=3.0in]{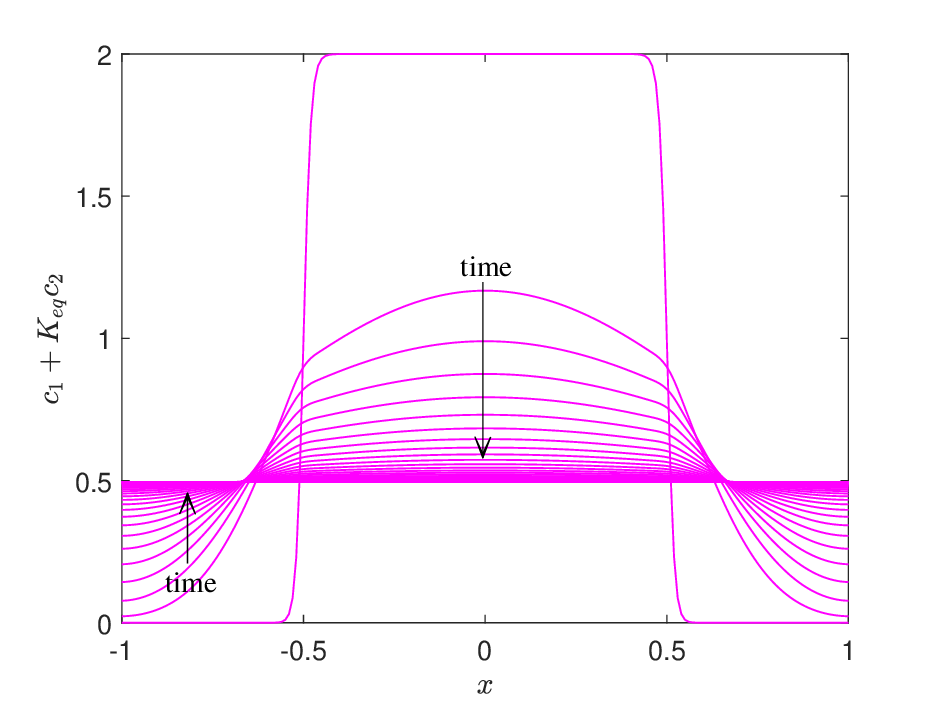}
		\end{minipage}
	}
	\caption{Evolutions of the concentrations $c_1$, $c_2$ and $c_1+K_{eq}c_2$ at different cases [(a) and (b): $D_1=D_2=K_{eq}=1$, (c) and (d): $D_1=D_2=1$ and $K_{eq}=1/3$, (e) and (f): $D_1=10$, $D_2=1$ and $K_{eq}=1/3$].}
	\label{fig_flat}
\end{figure}

Firstly, we consider the case of $D_1=D_2=K_{eq}=1$. The evolutions of concentrations $c_1$ and $c_2$ as well as $c_1+K_{eq}c_2$ are shown in Figs. \ref{fig_flat}(a) and \ref{fig_flat}(b), where the equilibrium jump of $c_1$ and $c_2$ is zero because of $K_{eq}=1$, and the final profile of $c_1+K_{eq}c_2$ becomes flat at the value of 1. 
Then we set $K_{eq}=1/3$ and plot the results in Figs. \ref{fig_flat}(c) and \ref{fig_flat}(d). From these two figures, one can observe that there is a concentration jump of $c_1$ and $c_2$ at the interface, but the final distribution of $c_1+K_{eq}c_2$ is flat at the equilibrium state.
Furthermore, the results in Figs. \ref{fig_flat}(e) and \ref{fig_flat}(f) demonstrate that the concentration profiles at equilibrium state are the same as those in Figs. \ref{fig_flat}(c) and \ref{fig_flat}(d), while the diffusion process presents some differences due to $D_1=10$. 
In conclusion, the concentrations change more rapidly with the increase of diffusivity $D_1$ or $D_2$, but the equilibrium distributions of them are determined by the concentration ratio $K_{eq}$. Additionally, it should be noted that all of the above results are consistent with those reported in Ref. \cite{Mirjalili2022A}.

\subsubsection{Linear equilibrium case}

We now continue to consider a droplet placed on the right half of the domain by $l\left(\mathbf{x}\right)=x$, the initial distributions of concentrations are $c_1\left(x,t=0\right)=2\phi$ and $c_2\left(x,t=0\right)=0$, and the Dirichlet boundary conditions are given by $c_1\left(x=0,t\right)=c_2\left(x=0,t\right)=0$ and $c_1\left(x=2,t\right)=2$, $c_2\left(x=2,t\right)=0$. In this case, the equilibrium solutions of concentrations are expected to be linear profiles with a jump at the interface. 

\begin{figure}	
	\centering
	\subfigure[]
	{
		\begin{minipage}{0.45\linewidth}		
			\centering		
			\includegraphics[width=3.0in]{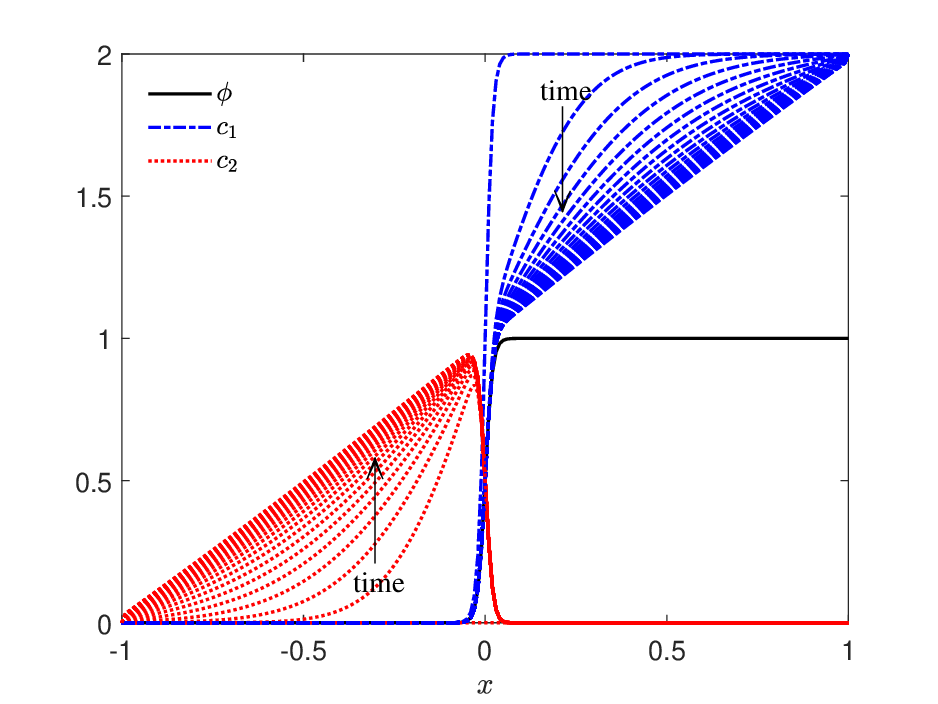}
		\end{minipage}
	}
	\subfigure[]
	{
		\begin{minipage}{0.45\linewidth}
			\centering
			\includegraphics[width=3.0in]{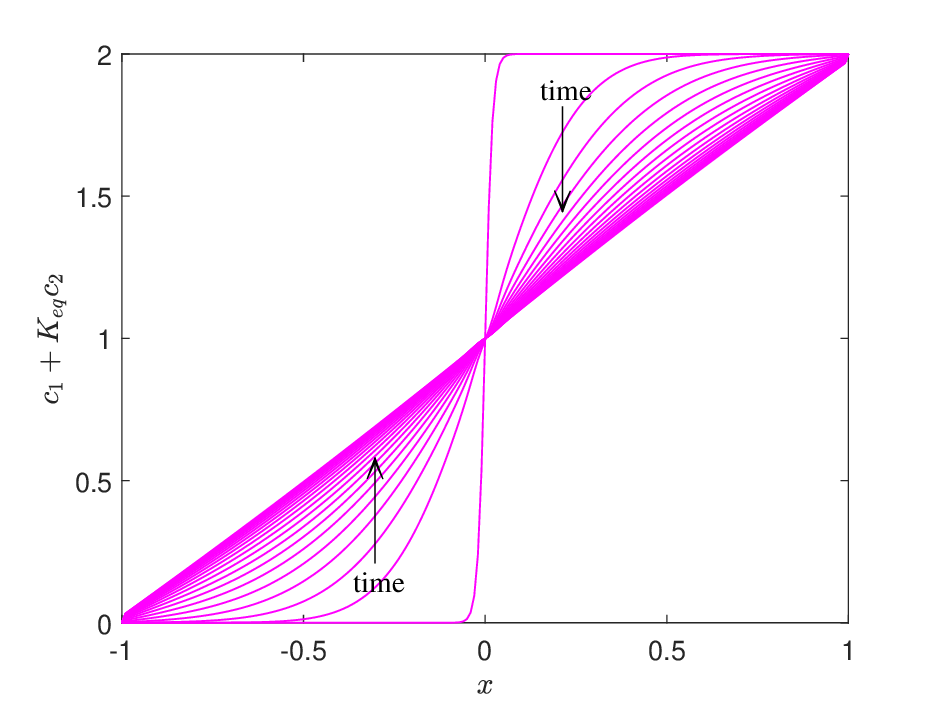}
		\end{minipage}
	}
	\subfigure[]
	{
		\begin{minipage}{0.45\linewidth}
			\centering
			\includegraphics[width=3.0in]{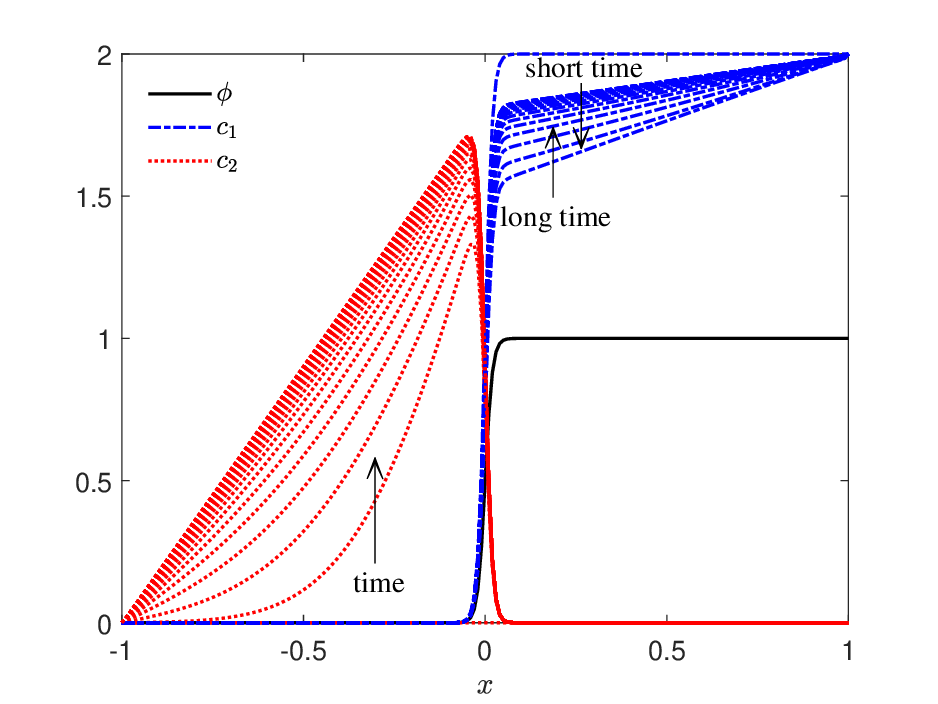}
		\end{minipage}
	}
	\subfigure[]
	{
		\begin{minipage}{0.45\linewidth}
			\centering
			\includegraphics[width=3.0in]{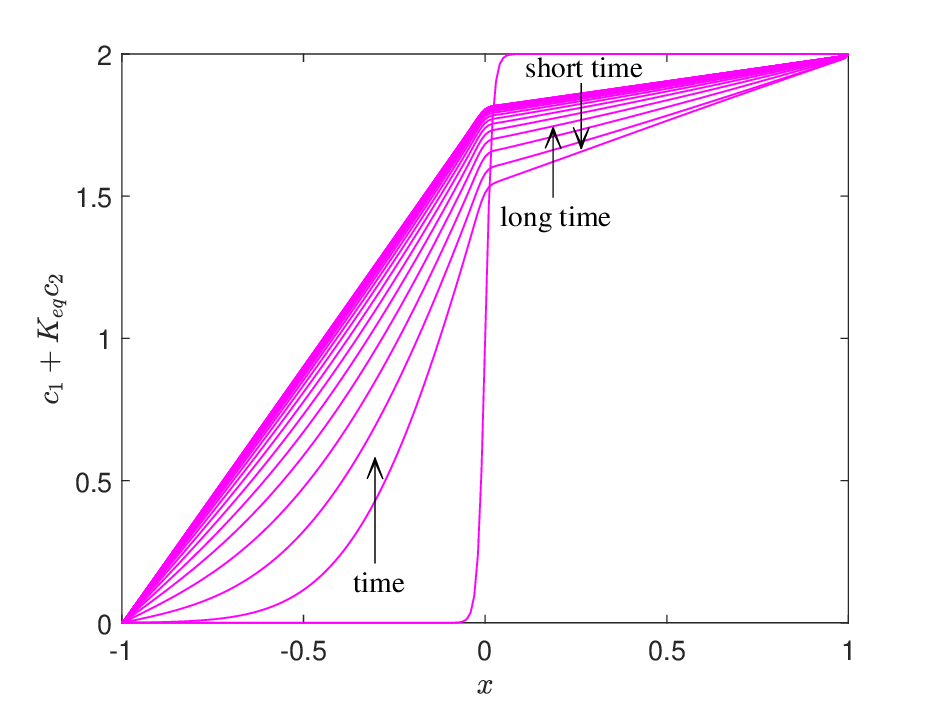}
		\end{minipage}
	}
	\subfigure[]
	{
		\begin{minipage}{0.45\linewidth}
			\centering
			\includegraphics[width=3.0in]{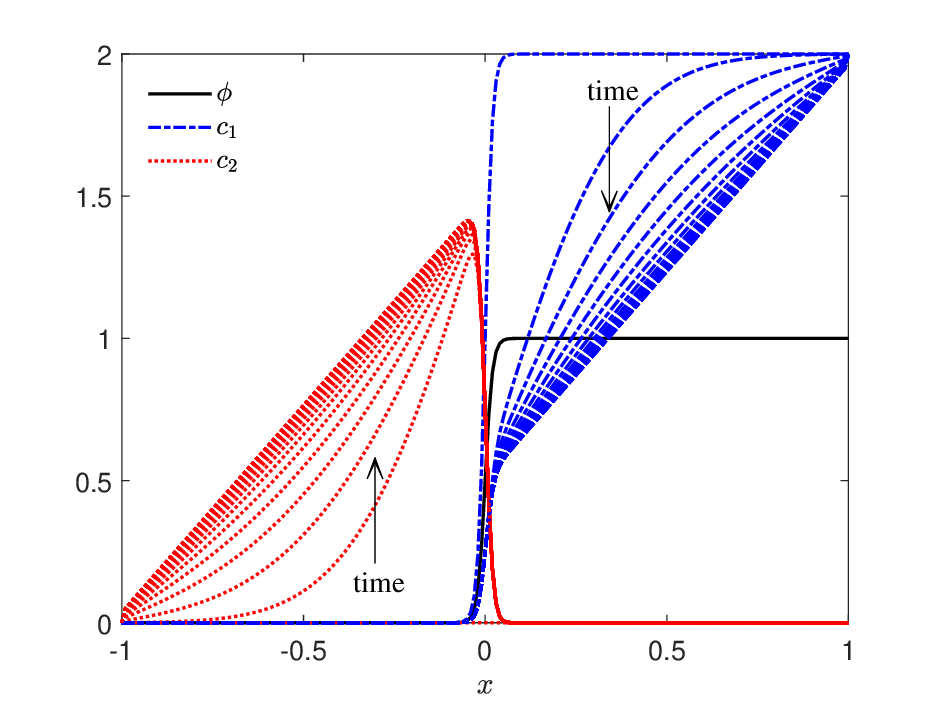}
		\end{minipage}
	}
	\subfigure[]
	{
		\begin{minipage}{0.45\linewidth}
			\centering
			\includegraphics[width=3.0in]{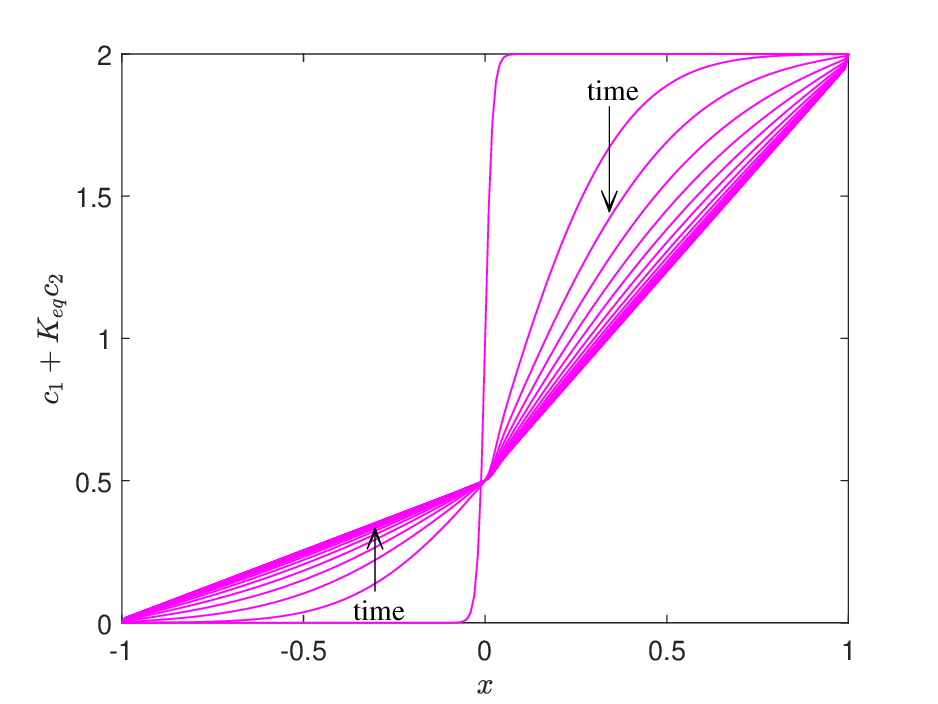}
		\end{minipage}
	}
	\caption{Evolutions of the concentrations $c_1$, $c_2$ and $c_1+K_{eq}c_2$ at linear equilibrium cases [(a) and (b): $D_1=D_2=K_{eq}=1$, (c) and (d): $D_1=10$, $D_2=1$ and $K_{eq}=1$, (e) and (f): $D_1=D_2=1$ and $K_{eq}=1/3$]. 
	}
	\label{fig_linear}
\end{figure}

We conduct some simulations with different parameters and present the results in Fig. \ref{fig_linear}. As shown in Figs. \ref{fig_linear}(a) where $D_1=D_2=K_{eq}=1$, the equilibrium concentrations at the interface remain continuous without any concentration jumps since $K_{eq}=1$, which is similar to that in Section \ref{flat_section}. Figure \ref{fig_linear}(b) shows the profile of $c_1+K_{eq}c_2$, and a linear equilibrium profile with a slope of 1 can be observed. 
When increasing the diffusivity in phase 1 to be $D_1=10$, the equilibrium concentration in phase 1 is expected to be greater than that in phase 2, which can be confirmed by the results in Figs. \ref{fig_linear}(c) and \ref{fig_linear}(d). In addition, there is also no concentration jump at the interface due to the choice of $K_{eq}=1$. Figures \ref{fig_linear}(e) and \ref{fig_linear}(f) present the concentration profiles under the condition of $K_{eq}=1/3$ and $D_1=D_2=1$. From these figures, one can find that when the system reaches to the equilibrium state, the concentration jump can be observed at the phase interface, which indicates that the slope of the concentration $c_1+K_{eq}c_2$ is greater than 1 in phase 1 but less than 1 in phase 2. Overall, the concentration profiles at equilibrium state in Fig. \ref{fig_linear} are in good agreement with those in the previous work \cite{Mirjalili2022A}.

\subsubsection{The case with large diffusivity ratio}

\begin{figure}
	\centering
	\subfigure[]
	{
		\begin{minipage}{0.45\linewidth}		
			\centering		
			\includegraphics[width=3.0in]{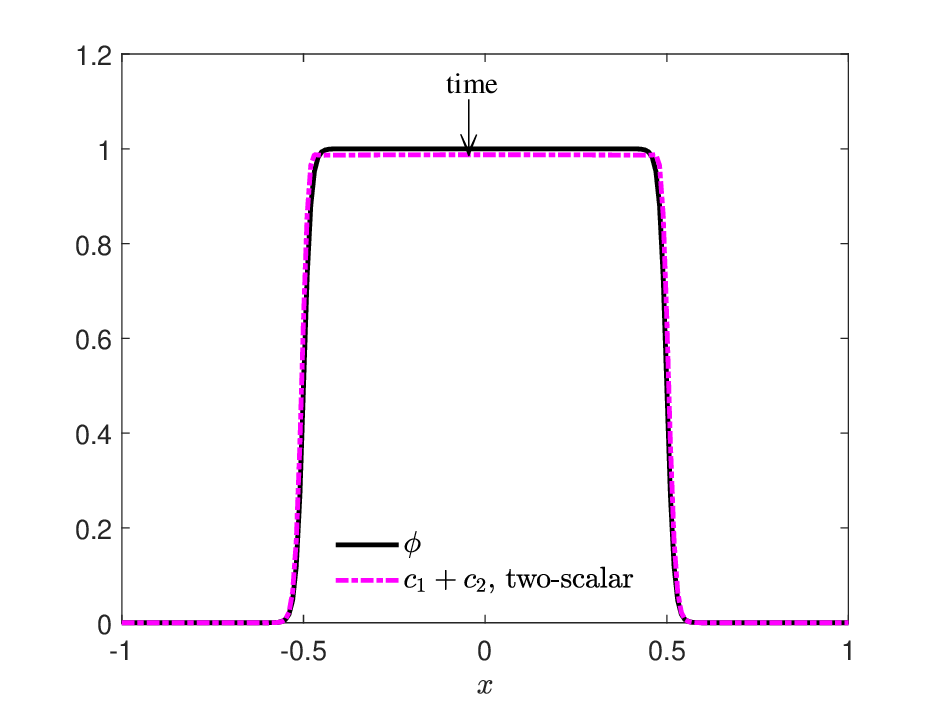}
		\end{minipage}
	}
	\subfigure[]
	{
		\begin{minipage}{0.45\linewidth}
			\centering
			\includegraphics[width=3.0in]{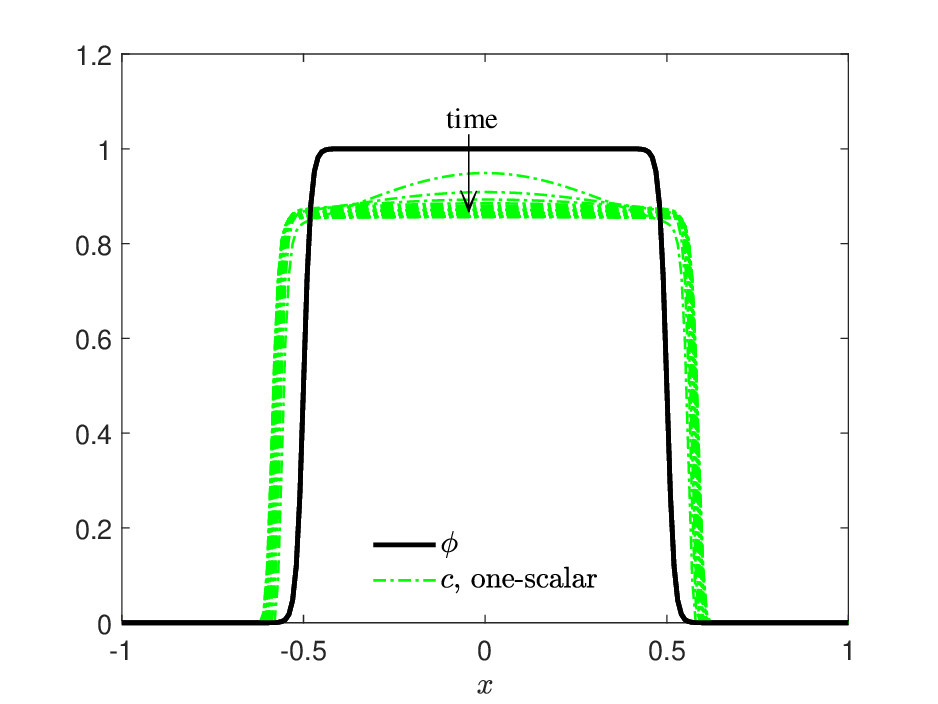}
		\end{minipage}
	}
	\caption{A comparison of one-scalar and two-scalar models for the case with large diffusivity ratio $D_1=1/D_2=10^{4}$ [(a) two-scalar model and (b) one-scalar model].}
	\label{fig_ratio}
\end{figure}

As highlighted in the previous work \cite{Mirjalili2022A}, the one-scalar model exhibits unphysical leakage for the problems with a large diffusivity ratio, whereas the two-scalar model can overcome this drawback. To confirm this statement, we consider an interfacial mass transfer problem with a large diffusivity ratio ($D_1=1$ and $D_2=10^{-4}$) and conduct some simulations. Initially, the concentrations are set as $c_1=\phi$ and $c_2=0$ in two-scalar model, whereas in the one-scalar model, $c=\phi$, and the order parameter is determined by Eq. (\ref{initial_phi}) with $l\left(\mathbf{x}\right)=-\left(x-0.5\right)\left(x+0.5\right)$. Figure \ref{fig_ratio} presents a comparison of the two models with $K_{eq}=1$. From this figure, one can see that the profile of concentration based on the two-scalar model is close to the initial distribution, while in the one-scalar model, there is a portion of the solute leaks from phase 1 to phase 2, which deviates a lot from the actual results.

\subsection{Multi-dimensional tests}
According to the results of one-dimensional numerical tests, the two-scalar model and the one-scalar model have a good agreement under the same diffusivity, and the two-scalar model can prevent the leakage of one-scalar model for the problem with a large diffusivity ratio. In the following, we will consider more practical problems to test present LB method.
\subsubsection{Bubble depletion}
\begin{figure}	
	\centering
	\subfigure[]
	{
		\begin{minipage}{0.45\linewidth}		
			\centering		
			\includegraphics[width=3.0in]{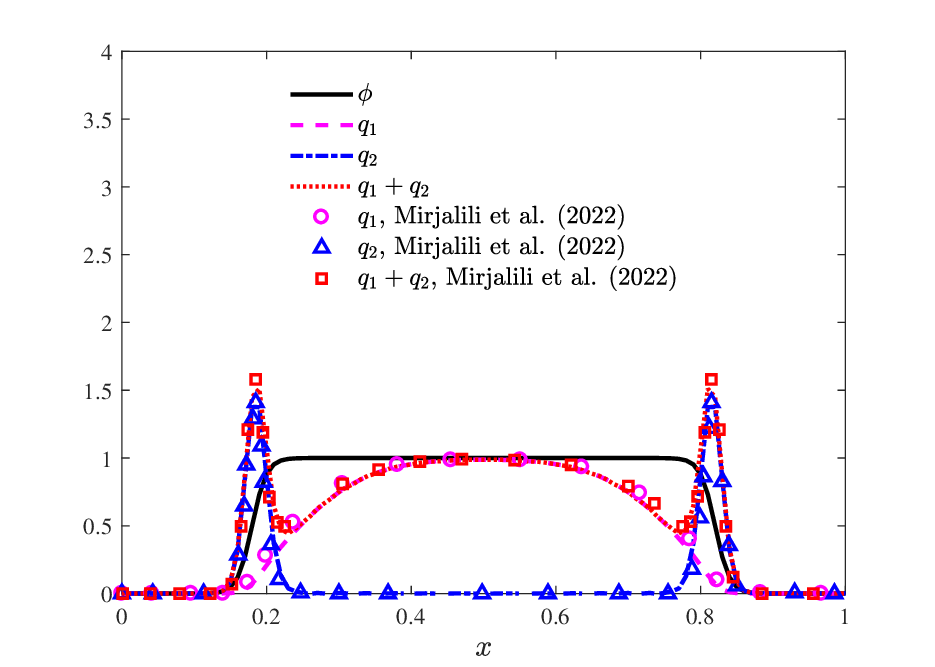}
		\end{minipage}
	}
	\subfigure[]
	{
		\begin{minipage}{0.45\linewidth}
			\centering
			\includegraphics[width=3.0in]{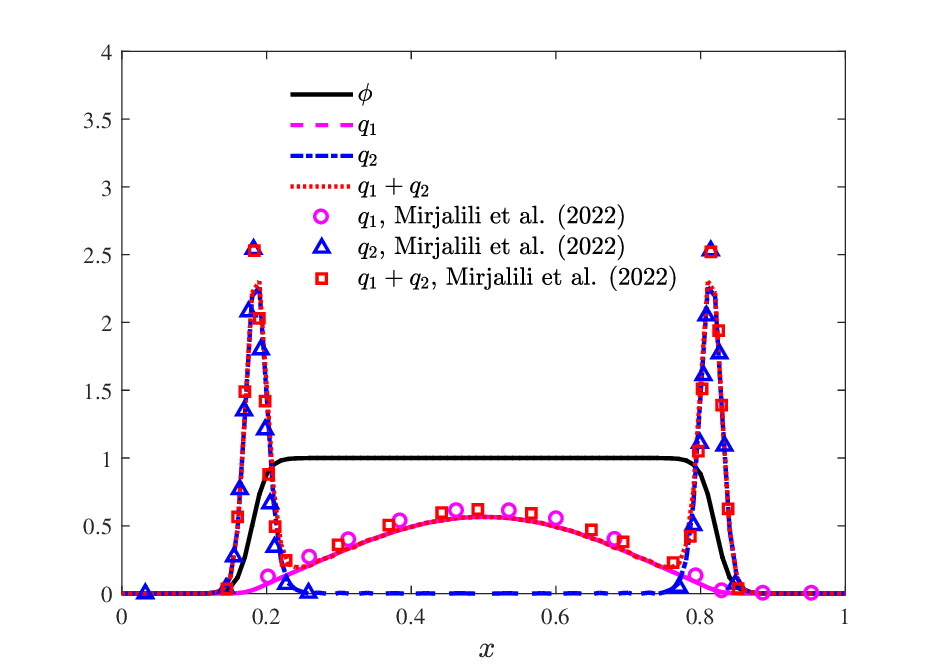}
		\end{minipage}
	}
	\subfigure[]
	{
		\begin{minipage}{0.45\linewidth}
			\centering
			\includegraphics[width=3.0in]{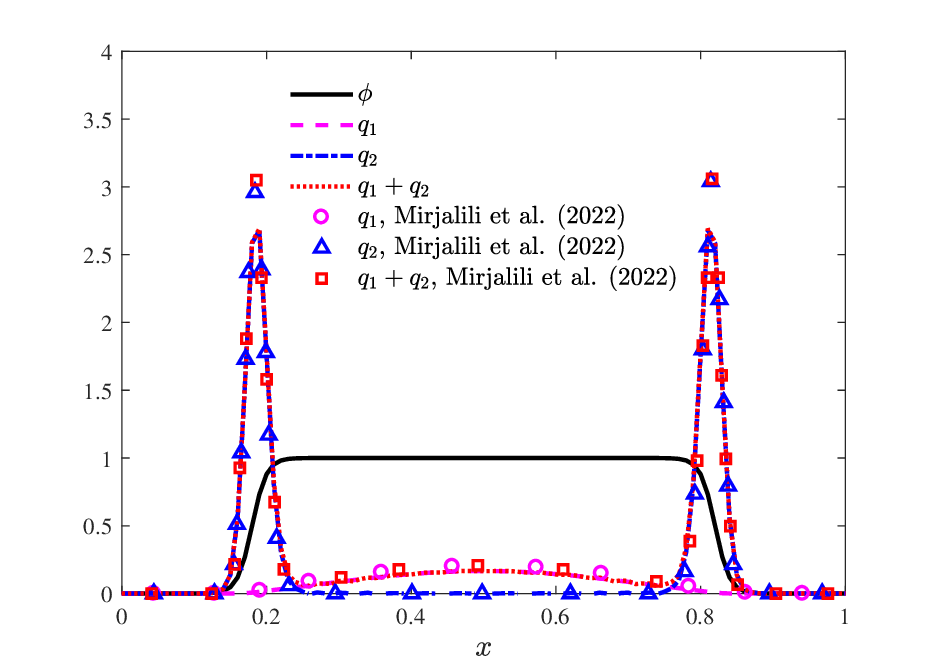}
		\end{minipage}
	}
	\subfigure[]
	{
		\begin{minipage}{0.45\linewidth}
			\centering
			\includegraphics[width=3.0in]{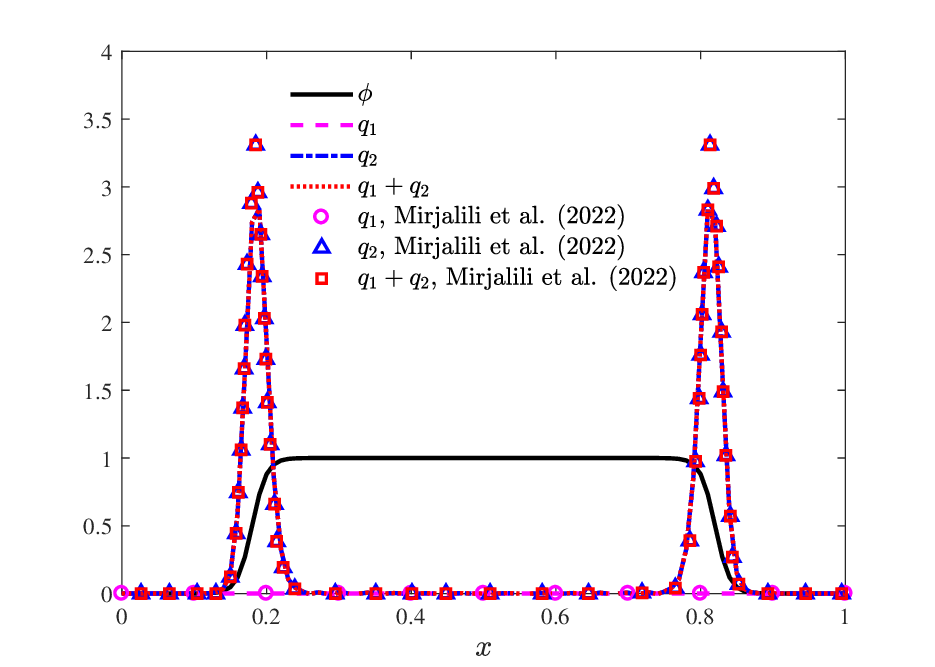}
		\end{minipage}
	}
	\caption{The predicted profiles of heat content at $y=z=0.5$ [(a) $t=13$, (b) $t=40$, (c) $t=80$, and (d) $t=250$].}
	\label{fig_bubble_dp}
\end{figure}

The problems considered above only involve the interfacial mass transfer between two phases, while in this part, the process of heat transfer in a three-dimensional bubble will be considered. A bubble (phase 1) with the radius $R=0.32$ is placed in the center of a cubic domain $1\times1\times1$, and it is surrounded by liquid of phase 2. Specifically, the signed-distance function for order parameter is defined by $l\left(\mathbf{x}\right)=R-\sqrt{{(x-0.5)}^2+{(y-0.5)}^2+{(z-0.5)}^2}$. The heat content is initialized to be $q_1=\phi$, $q_2=0$, and they will diffuse into the whole domain until reach to the equilibrium state under the physical parameters of $D_1=3.62\times 10^{-4}$, $D_2=3.62\times 10^{-9}$, and $K_{eq}=10^{-5}$. In our simulations, the grid size is $100\times100\times100$ and time step is $\delta t=10^{-4}$. Figure \ref{fig_bubble_dp} shows the heat contents solutions of two-scalar model at $y=z=0.5$ and different moments. From this figure, we can observe that the heat diffusing outward from the bubble is easy to accumulate at the gas-liquid interface.
Additionally, it can be seen from the figure that our results are in good agreement with previous results \cite{Mirjalili2022A}, and also, the total heat content of the system $q_1+q_2$ is always conserved during the evolution, which also indicates the accuracy of the LB method. 

\par
\begin{figure}	
	\centering
	\includegraphics[width=3.5in]{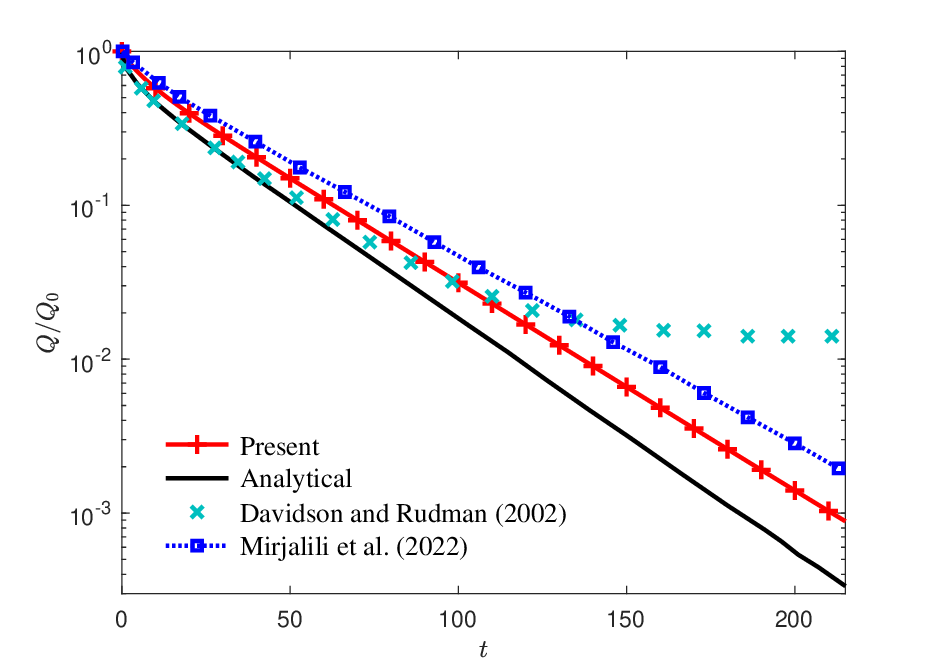}
	\caption{Evolution of the normalized heat content in the gas phase.}
	\label{fig_bubble_dt}
\end{figure}

 To further make a quantitative comparison, we calculate the normalized total heat content in the gas phase $Q/Q_0$, where $Q_0=\int q_1\, dV$ is the initial total heat content. As seen from Fig. \ref{fig_bubble_dt}, the present result is close to the analytical solution and is more accurate than the data in Refs. \cite{Mirjalili2022A, Davidson_Rudman_2002}. 

\subsubsection{Rising bubble}
\begin{figure}	
	\centering
	\subfigure[]
	{
		\begin{minipage}{0.3\linewidth}		
			\centering		
			\includegraphics[width=2.0in]{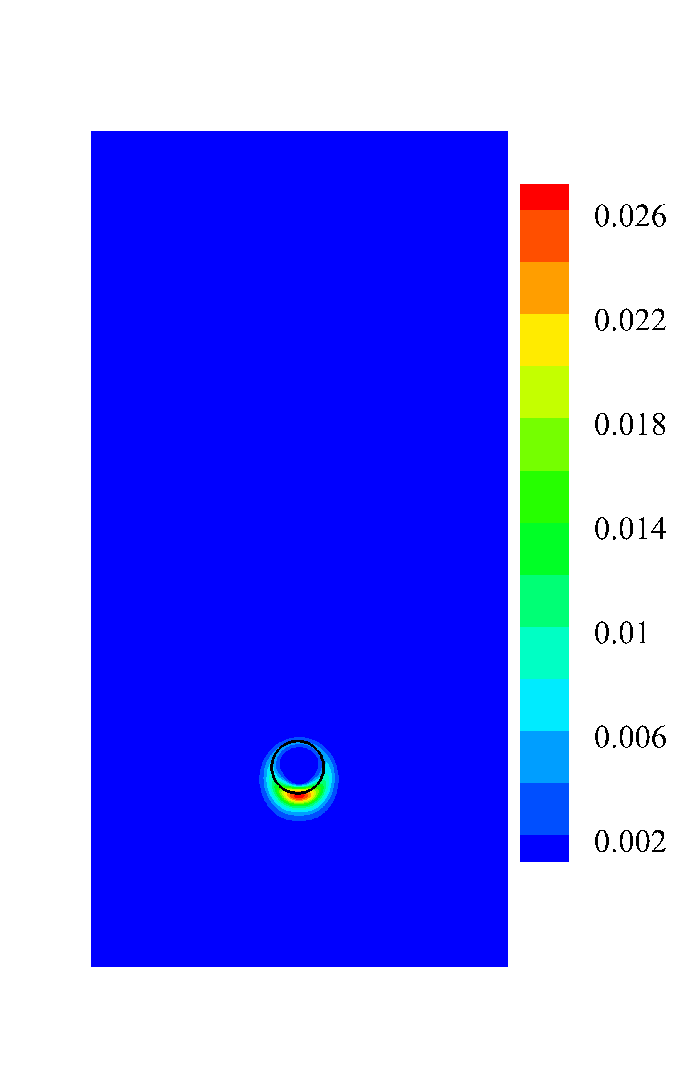}
		\end{minipage}
	}
	\subfigure[]
	{
		\begin{minipage}{0.3\linewidth}
			\centering
			\includegraphics[width=2.0in]{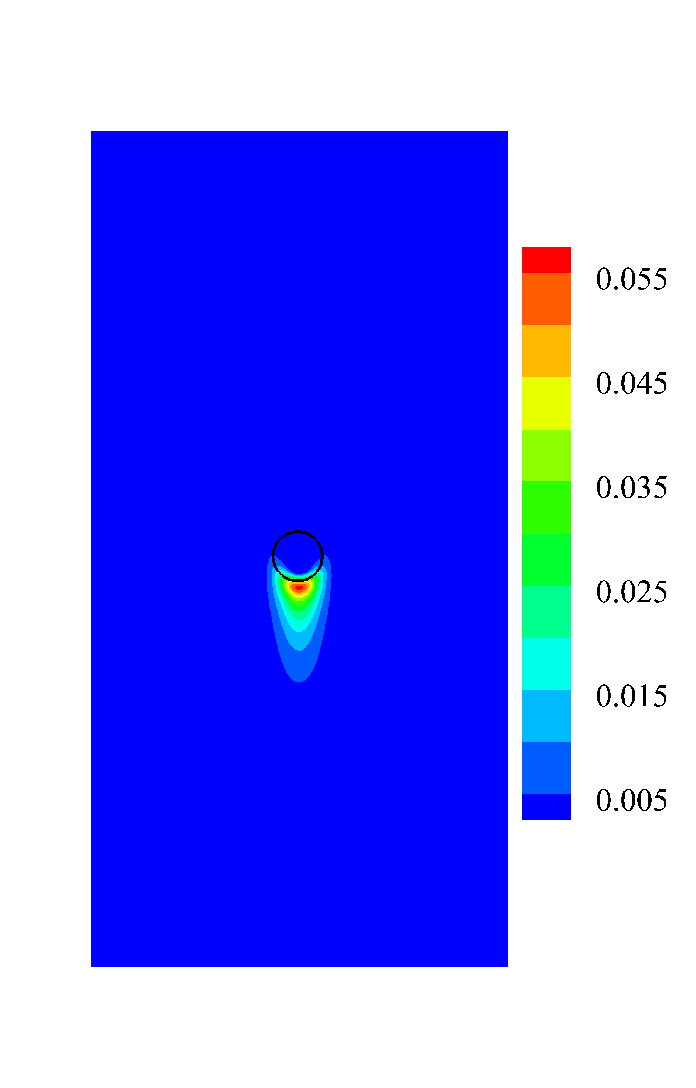}
		\end{minipage}
	}
	\subfigure[]
	{
		\begin{minipage}{0.3\linewidth}
			\centering
			\includegraphics[width=2.0in]{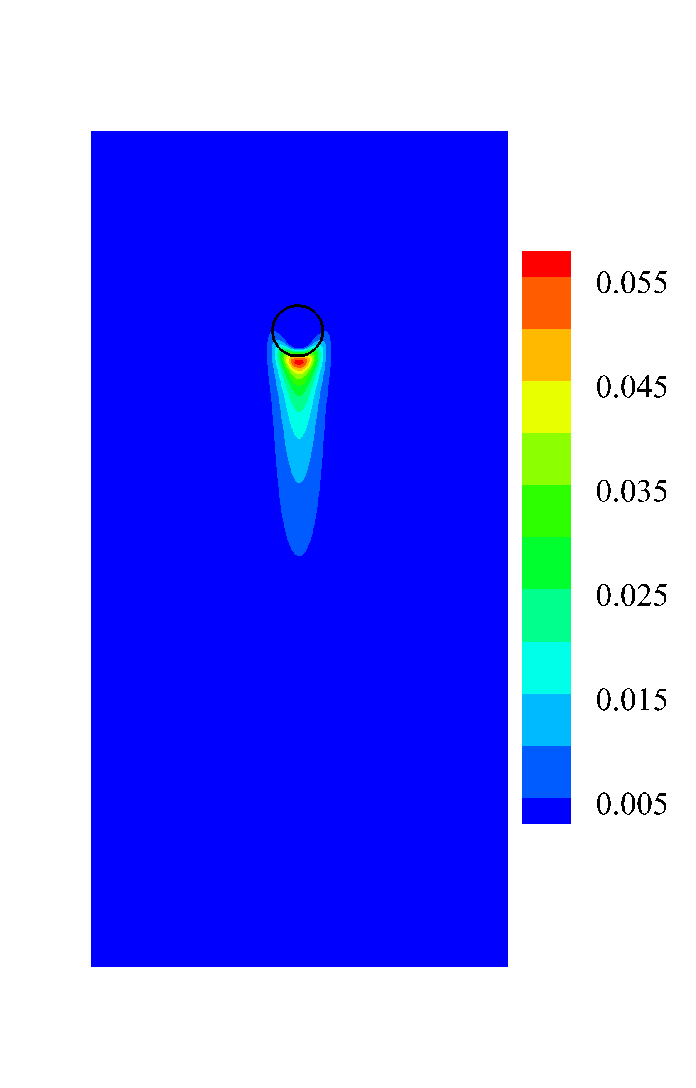}
		\end{minipage}
	}
	\caption{Snapshots of the oxygen concentration during bubble rising [(a) $t=0.21\,\si{s}$, (b) $t=0.84\,\si{s}$, (c) $t=1.47\,\si{s}$, the solid black line represents the interface of two phases].}
	\label{fig_bubble_t}
\end{figure}

It is noted that in above problems, only phase field and mass/heat transfer are involved. To further validate the present LB method, the influence of fluid field is also considered. To this end, the interfacial transfer of oxygen content within a two-dimensional rising bubble is investigated. In our simulations, a gas bubble (phase 2) with the radius $R=0.5\,\si{mm}$ is initially located at the position $(4\,\si{mm}, 3\,\si{mm})$ in the square computational domain $8\,\si{mm} \times 16\,\si{mm}$, and thus the signed-distance function is given by $l\left(\mathbf x\right)=\sqrt{{(x-0.0004)}^2+{(y-0.0003)}^2}-R$. The oxygen contents in gas and liquid phases are initialized by $c_1=0$ and $c_2=1-\phi$. In this problem, we consider the material properties of two fluids to be $\rho_1=1000\, \si{kg/m^3}$, $\rho_2=1.2\, \si{kg/m^3}$, $\mu_1=1\times 10^{-2} \,\si{Pa\, s}$, $\mu_2=1.8\times 10^{-5} \,\si{Pa\, s}$, $\sigma=0.0728\,\si{kg/s^2}$, and the buoyancy force with $g=9.8\,\si{m/s^2}$ is imposed on the two fluids. The diffusivities of oxygen in gas and liquid are $D_1=10^{-6} \,\si{m^2/s}$ and $D_2=5\times10^{-6} \,\si{m^2/s}$, and the Henry's coefficient is $K_{eq}=1/33$ at standard conditions. Additionally, the grid size is set to be $192\times384$, and the interface thickness is $W=5\Delta x$. Figure \ref{fig_bubble_t} plots the oxygen content in water ($c_2$) at the moments of $t=0.21\,\si{s}, 0.84\,\si{s}, 1.47\,\si{s}$. From this figure, it can be found that with the rise of the bubble, the oxygen content around the bubble gradually changes from a ring shape to a dragged wake, but, the maximum value of the oxygen content is always near the bottom of the bubble.

\begin{figure}	
	\centering
	\includegraphics[width=3.5in]{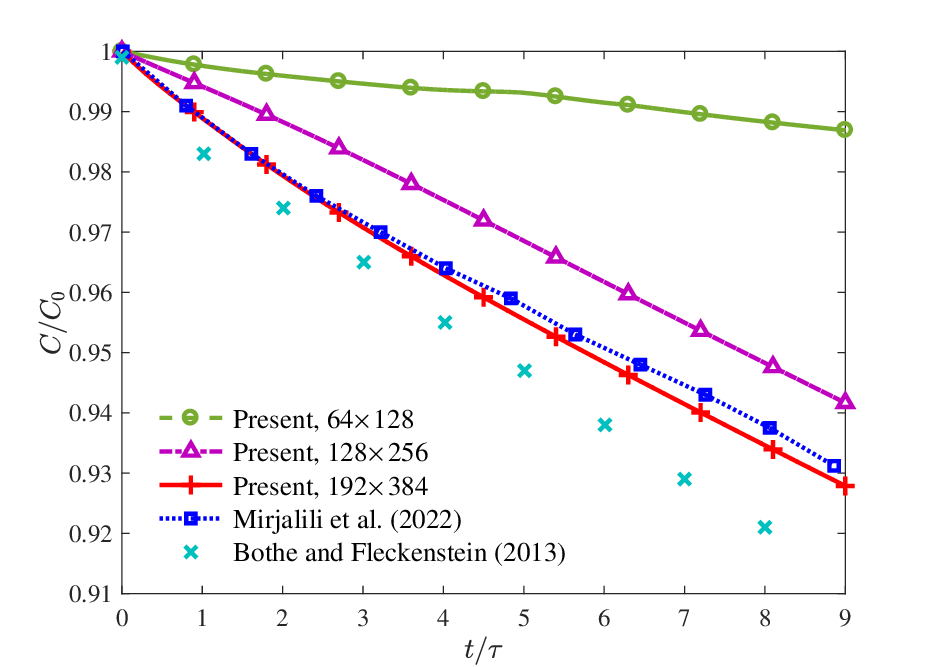}
	\caption{Evolution of the normalized oxygen content in the gas phase.}
	\label{fig_bubble_r}
\end{figure}

A further analysis of the rising bubble is shown in Fig. \ref{fig_bubble_r} where the normalized total oxygen content in water $C/C_0$ under three different grid resolutions are displayed, and a larger grid resolution can lead to the better convergence of the model predictions. In this figure, $C_0=\int c_2\, dV$ is the initial total heat content, $\tau=0.013\,\si{s}$ is a dimensionless time, and one can find that our results with a larger grid resolution are more close to the reference \cite{Bothe_Fleckenstein_2013}. 

\subsection{Concentration/temperature-dependent viscosity system}
We note that the viscosity in two-phase system with the mass/heat transfer is usually given by Eq. (\ref{1.8}), which is the constant value in pure fluids. However, this relation may be not ture since the viscosity should also depend on the concentration and temperature \cite{grunberg1949mixture,viscosity-concentration1, viscosity-concentration2}. In general, the increase of concentration leads to an increase of the fluid viscosity, and in the following, the exponential form is adopted,
\begin{equation}\label{4.3}
	\log \,\mu=c_1\log\, \mu_1+c_2\log \,\mu_2.
\end{equation}

\subsubsection{Layered Poiseuille flow coupled with concentration field}

We first focus a layered two-phase flow driven by a constant force $\mathbf{G}=(G_x,0)$ in a two-dimensional channel $[-0.05, 0.05]\times[-0.5, 0.5]$. The fluid interface is described by the signed-distance function $l\left(\mathbf{x}\right)=-y$, the initial concentrations are set to be $c_1=\phi$ and $c_2=0$, and the diffusivities are $D_1=D_2=1$. In our simulations, the grid size is $15\times150$, the densities $\rho_1=10$ and $\rho_2=1$, and the viscosities $\mu_1=1$ and $\mu_2=0.1$.

\begin{figure}	
	\centering
	\subfigure[]
	{
		\begin{minipage}{0.45\linewidth}		
			\centering		
			\includegraphics[width=3.0in]{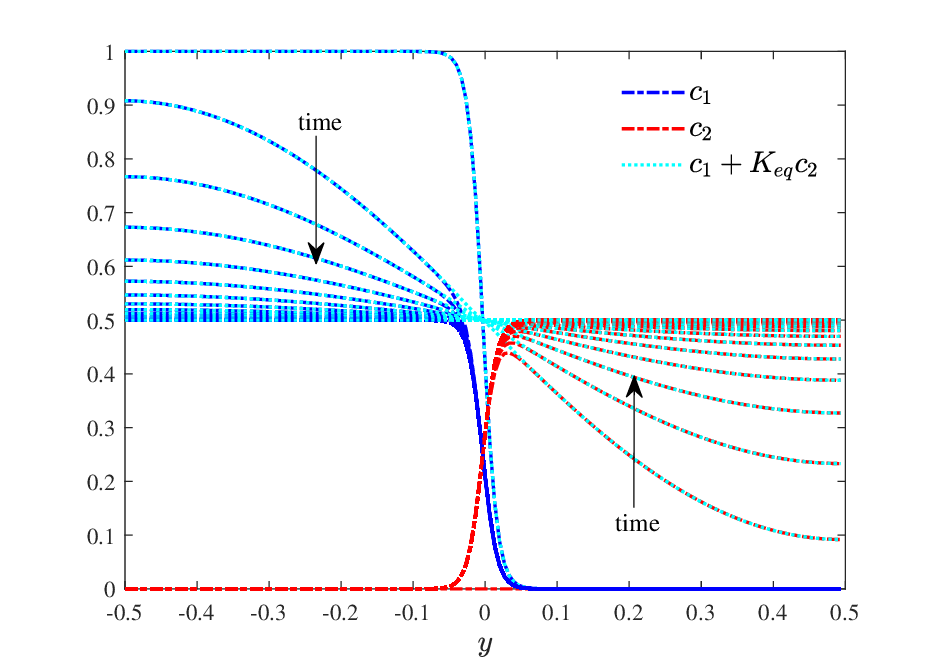}
		\end{minipage}
	}
	\subfigure[]
	{
		\begin{minipage}{0.45\linewidth}
			\centering
			\includegraphics[width=3.0in]{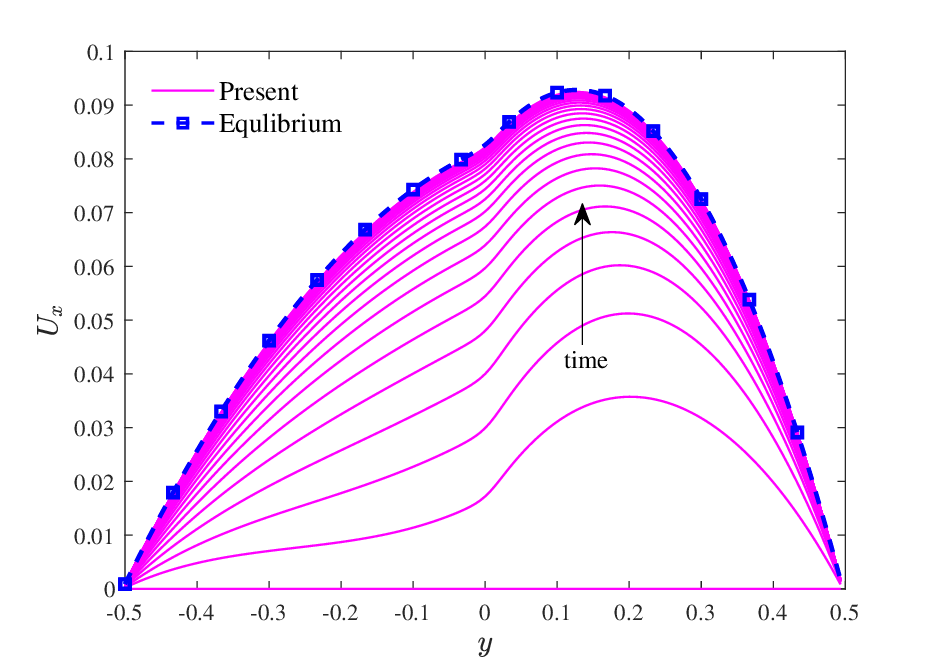}
		\end{minipage}
	}
	\caption{Evolutions of some physical quantities of the layered Poiseuille flow [(a) concentrations $c_1$, $c_2$, and $c_1+K_{eq}c_2$, (b) horizontal velocity].}
	\label{fig_uxuy}
\end{figure}

For the first case with $K_{eq}=1$, as shown in Fig. \ref{fig_uxuy}(a), the concentration distribution reaches to a flat profile at the steady state. According to the equilibrium concentration $c_1=c_2=0.5$ and Eq. (\ref{4.3}), we can obtain the analytical solution of the horizontal velocity
\begin{equation}
\log \,\mu=0.5\log\, \mu_1+0.5\log \,\mu_2.
\end{equation}
Figure \ref{fig_uxuy}(b) shows the evolution of the numerical horizontal velocity, which is also consistent with the above analytical solution at the final steady state.

\begin{figure}	
	\centering
	\subfigure[]
	{
		\begin{minipage}{0.4\linewidth}		
			\centering		
			\includegraphics[width=2.8in]{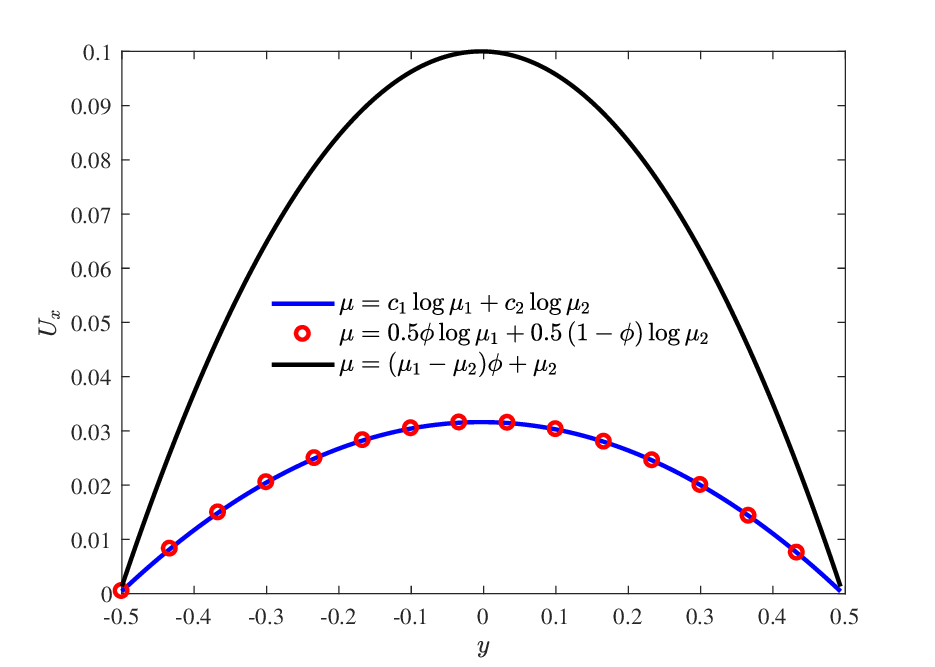}
		\end{minipage}
	}
	\subfigure[]
	{
		\begin{minipage}{0.4\linewidth}		
			\centering		
			\includegraphics[width=2.8in]{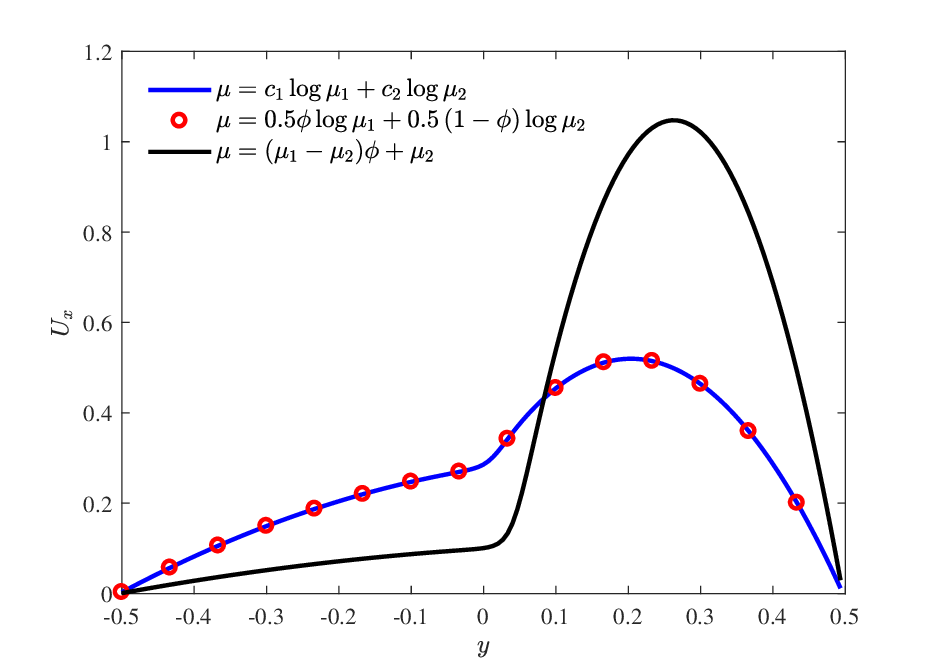}
		\end{minipage}
	}
	\subfigure[]
	{
		\begin{minipage}{0.4\linewidth}		
			\centering		
			\includegraphics[width=2.8in]{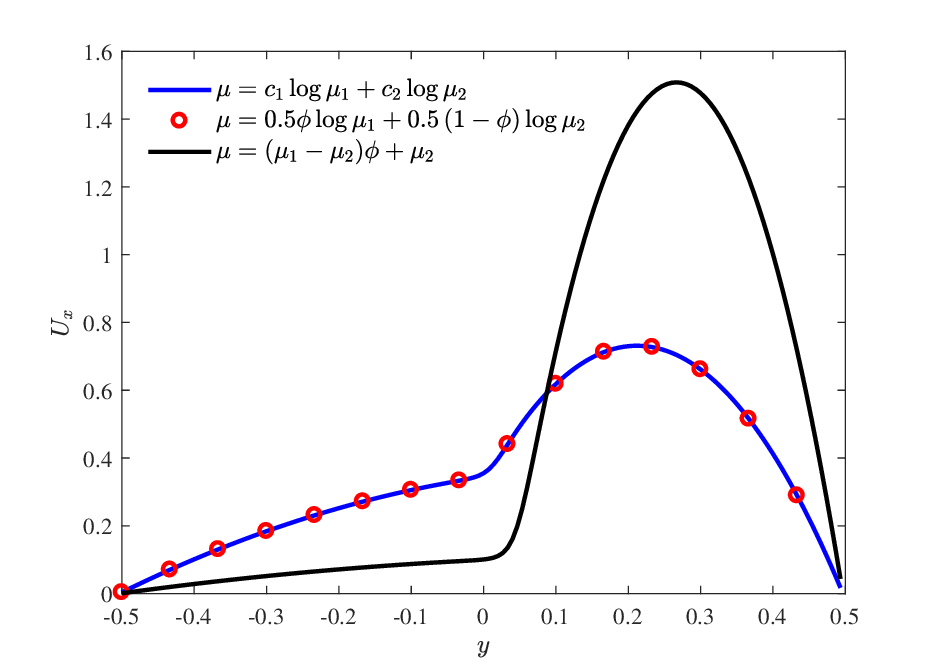}
		\end{minipage}
	}
	\subfigure[]
	{
		\begin{minipage}{0.4\linewidth}
			\centering
			\includegraphics[width=2.8in]{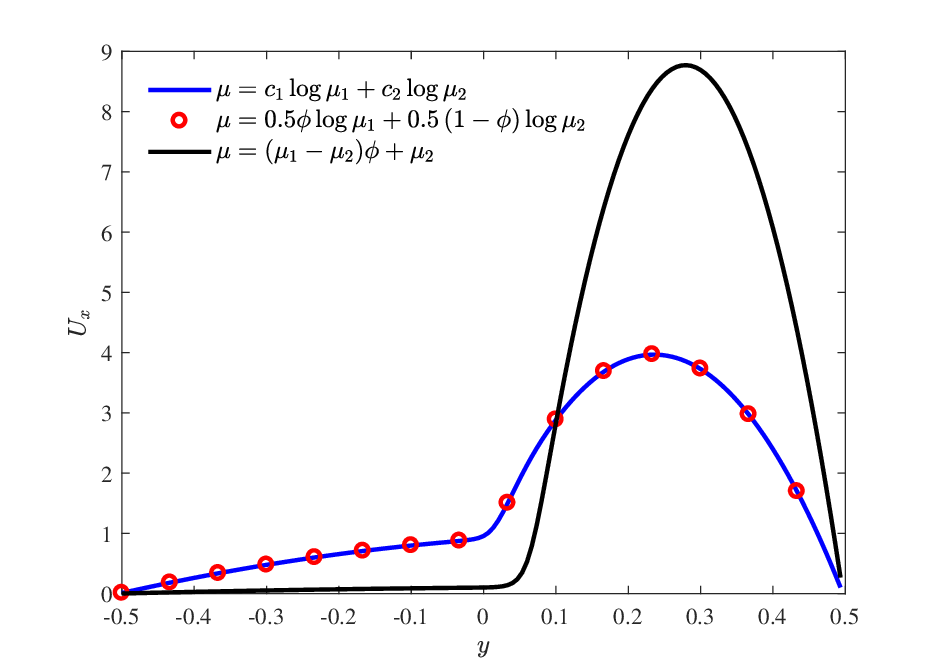}
		\end{minipage}
	}
	\caption{The predicted horizontal velocities under different density ratios [(a) $\rho_1=1$, (b) $\rho_1=100$, (c) $\rho_1=150$, (d) $\rho_1=1000$].}
	\label{fig_rho1}
\end{figure}

To show the effect of concentration field on the fluid velocity, we conduct a comparison between the concentration-independent viscosity and the concentration-dependent viscosity. As seen from Fig. \ref{fig_rho1}, the maximum velocity of the flow field decreases when the influence of concentration is included, which can be explained by the fact that the increase of concentration gives rise to a higher viscosity of the fluid, and the maximum velocity would be reduced when other conditions are not changed.

\begin{figure}	
	\centering
	\includegraphics[width=3.5in]{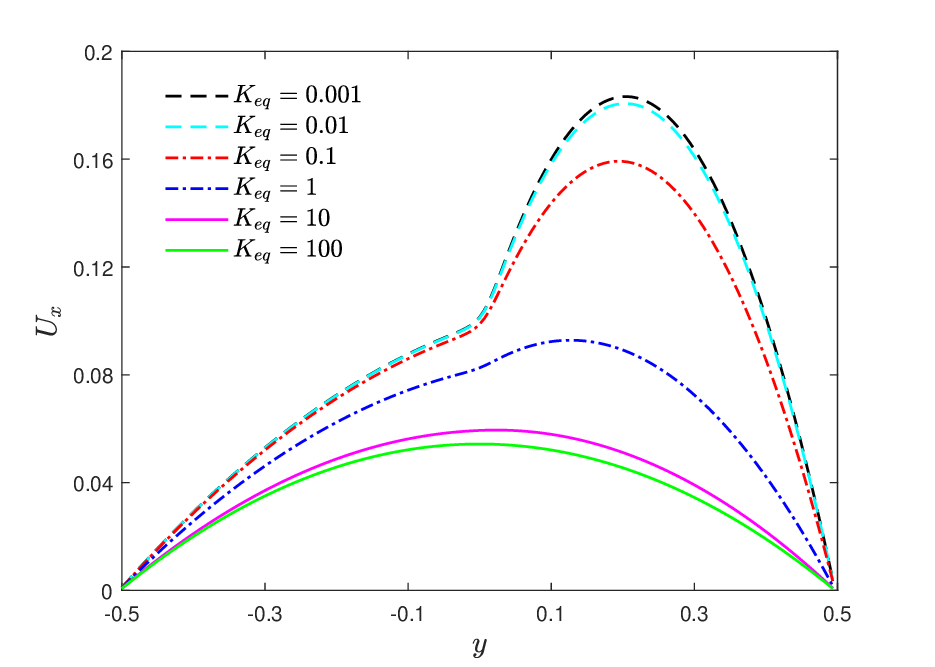}
	\caption{The predicted horizontal velocity with different values of $K_{eq}$.}
	\label{fig_keq1}
\end{figure}

Finally, we consider the effect of interfacial concentration ratio with $K_{eq}$ varying from $0.001$ to $100$, and plot the results in Fig. \ref{fig_keq1} where $\rho_1=10$. From this figure, one can observe that with the increase of $K_{eq}$, the horizontal velocity of the flow field gradually decreases, which can be attributed to the fact that when the initial concentrations in two phases are the same, an increase in the concentration ratio will lead to a larger concentration gradient between the two phases, which results in an increase of flow resistance and a decrease of the horizontal velocity. 

\subsubsection{Gas displacement of crude oil}

\begin{figure}
	\centering
	\subfigure[]
	{
		\begin{minipage}{0.3\linewidth}		
			\centering		
			\includegraphics[width=2.0in]{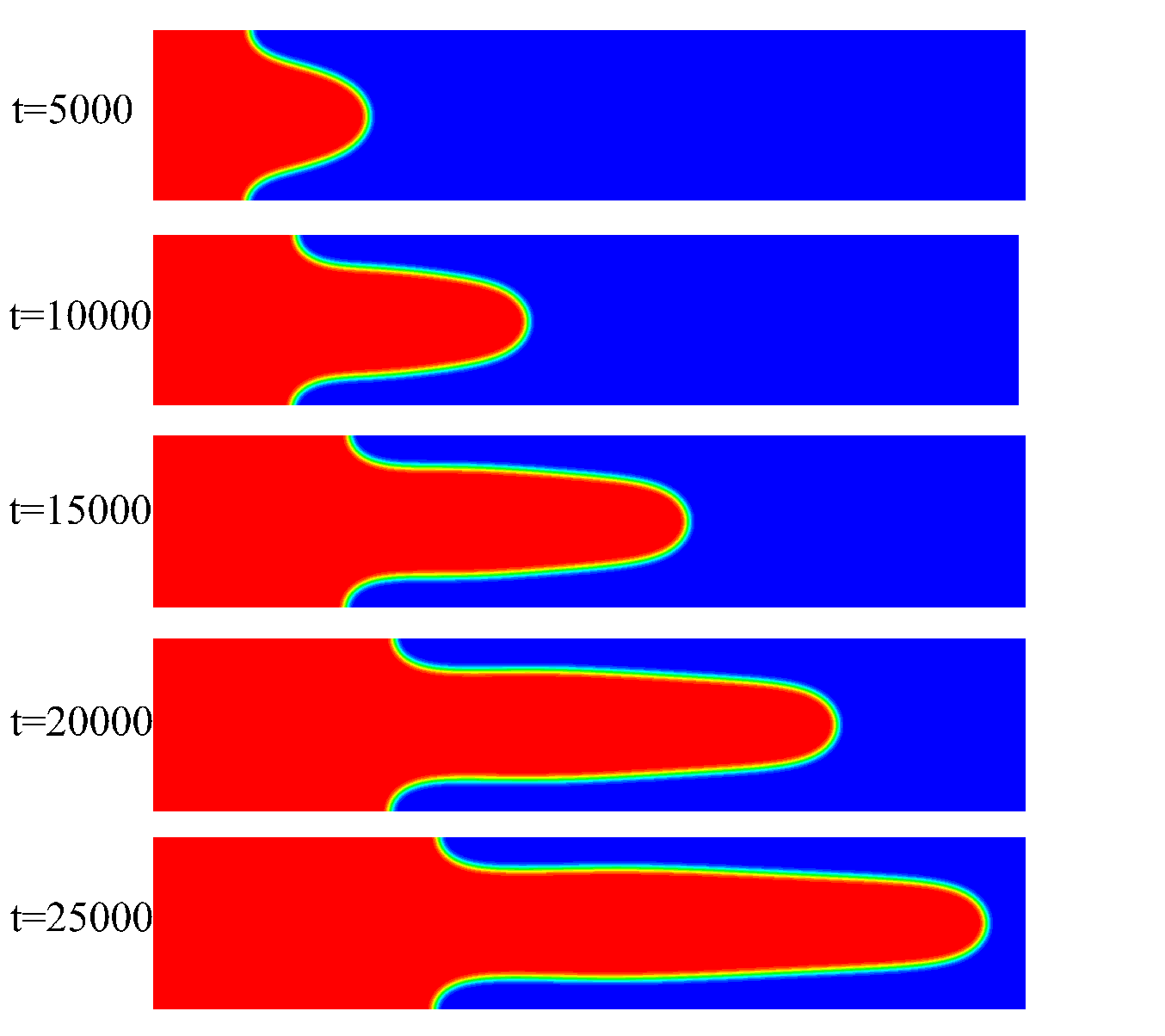}
		\end{minipage}
	}
	\subfigure[]
	{
		\begin{minipage}{0.3\linewidth}		
			\centering		
			\includegraphics[width=2.0in]{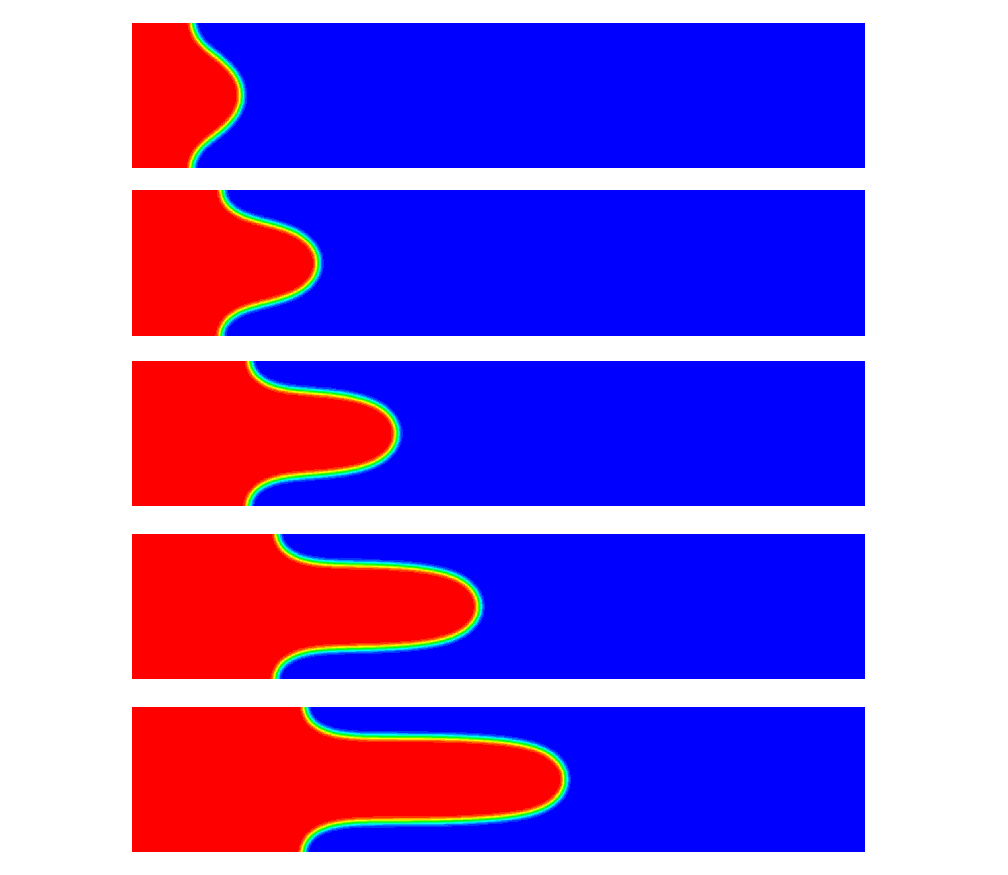}
		\end{minipage}
	}
	\subfigure[]
	{
		\begin{minipage}{0.3\linewidth}
			\centering
			\includegraphics[width=2.0in]{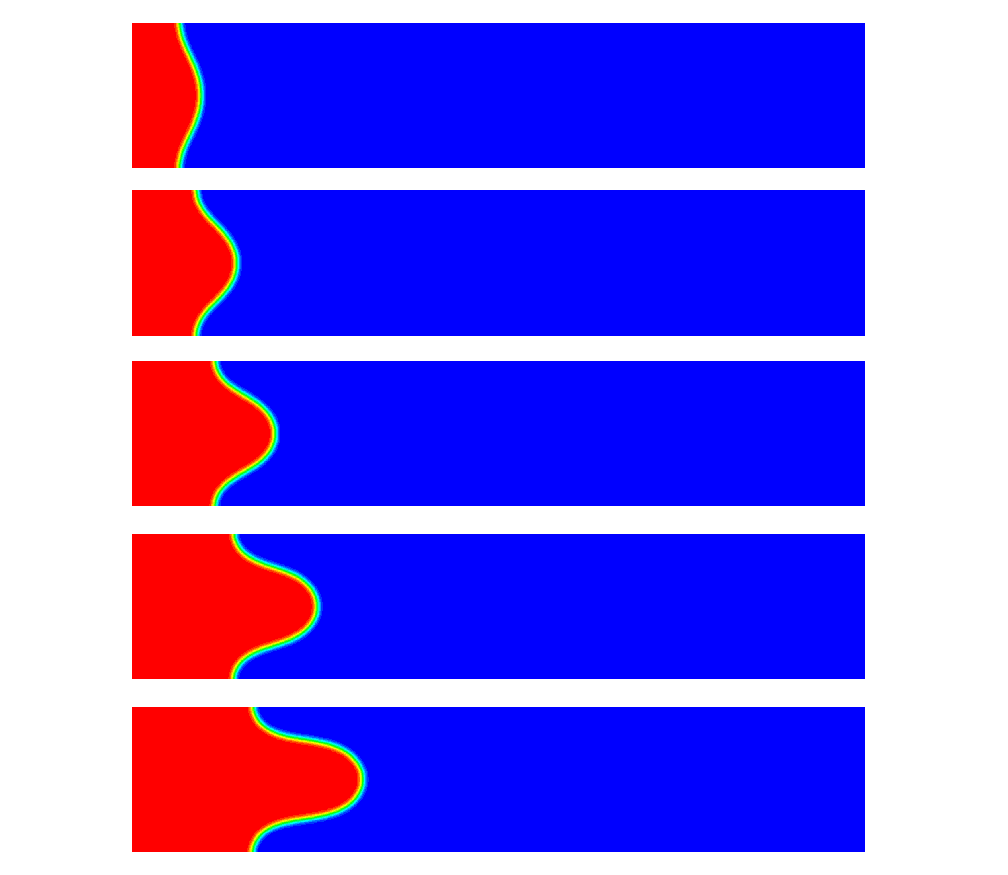}
		\end{minipage}
	}
	\caption{The displacement of crude oil by gas under different initial concentrations in oil [(a) $c_2=0.001$, (b) $c_2=0.4$, (c) $c_2=0.8$].}
	\label{fig_disp}
\end{figure}

The last problem we considered is the displacement of liquid by gas under different saturations. A channel with length $L=4.0$ and width $W=0.8$ is full of the liquid phase except for the region $x\textless0.1$ filled by gas (phase 1), i.e., $l\left(\mathbf x\right)=0.1-x$. At the beginning, the concentration in the gas is fixed as $c_1=0.8$, and other physical parameters are $D_1=D_2=1$ and $K_{eq}=2$. The properties of two pure fluids are set as $\rho_1=\rho_2=1$, $\mu_{1}=1$, $\mu_{2}=10$, and $\sigma=2\times 10^{-9}$. The fluids are driven by the pressure drop $\Delta p=0.01$. In this study, the lattice spacing is $\Delta x=0.01$ and time step is $\delta t=10^{-4}$. Figure \ref{fig_disp} presents the displacement process of two-phase fluids with different initial concentrations $c_2$ at the same time interval. From this figure, it can be seen that when the initial concentration in liquid phase is small [see Fig. \ref{fig_disp}(a)], the concentration gradient between the two phases is large, and the displacement speed is the fastest. When the concentration in liquid phase increases to $0.4$ and $0.8$, as shown in Figs. \ref{fig_disp}(b) and \ref{fig_disp}(c), the displacement velocity gradually decreases with the decrease of concentration difference between the two phases. These results also demonstrate that the concentration field has a significant influence on the velocity of viscous fingering.

\section{Conclusions}\label{Conclusion}

In this paper, we develop a new LB method for the conjugate interfacial mass/heat transport in two-phase flows. Through the Chapman–Enskog analysis, the governing equations in Ref. \cite{Mirjalili2022A} can be correctly recovered from the proposed LB method. In the present LB method, some proper auxiliary terms in LB evolution equations are constructed to recover the complex nonlinear source terms in the two-scalar model for mass/heat transfer, and the discretizations of some derivative terms can be avoided. In addition, the LB model for one-scalar equation of concentration field is also developed to perform some comparisons. To test the present LB method and show the difference between one-scalar and two-scalar models, we first consider some one-dimensional problems, and the numerical results are in good agreement with the analytical solutions and data reported in some previous works. To conduct more validations, some practical multi-dimensional problems are  also studied, and the results illustrate the accuracy and efficiency of present LB method. Moreover, the systems with the concentration/temperature-dependent viscosity are also considered, and the results show that the distribution of concentration has a significant influence on the flow field.

\begin{acknowledgments}
	This research has been supported by the National Natural Science Foundation of China (Grants No. 12072127 and No. 51836003), the Interdiciplinary Research Program of HUST (No. 2023JCYJ002), and the Fundamental Research Funds for the Central Universities, HUST (No. 2023JYCXJJ046). The computation was completed on the HPC Platform of Huazhong University of Science and Technology.
\end{acknowledgments}   

\appendix*

\section{The Chapman-Enskog analysis of the lattice Boltzmann model for the two-scalar equations}\label{CE_analysis}
For simplicity but without losing the generality, we only conduct the Chapman-Enskog analysis on the LB model for Eq. (\ref{two-scalar_1}). We first expand the distribution functions $h_{1,i}$, $H_{1,i}$, $\widetilde H_{1,i}$, as well as the time and space derivatives in different order of a small expansion parameter $\epsilon$ as
\begin{equation}
	\begin{aligned}\label{expansion}
		h_{1,i}=h_{1,i}^{(0)}+\epsilon h_{1,i}^{(1)}+{\epsilon^2}h_{1,i}^{(2)}+\cdots,\\
		H_{1,i}=\epsilon H_{1,i}^{(1)}+\epsilon^2 H_{1,i}^{(2)}, \widetilde H_{1,i}={\epsilon^2}\widetilde H_{1,i}^{(2)}, \\
		\partial_t=\epsilon \partial_{t_1}+\epsilon^2 \partial_{t_2}, \nabla=\epsilon \nabla_1.
	\end{aligned}
\end{equation}
According to Eqs. (\ref{h_equilibrium}) and (\ref{macro}), one can obtain the following moment conditions:
\begin{equation}
	\begin{aligned}
		\sum_i h_{1,i}^{eq}=c_1, \sum_i\mathbf c_i h_{1,i}^{eq}&=c_1\mathbf{u}, \sum_i\mathbf c_i \mathbf c_i h_{1,i}^{eq}=c_s^2 c_1\mathbf I, \sum_i {H_{1,i}}={M_{H_{1,i}}}=AD_m \left[K_{eq}c_2\phi-c_1(1-\phi)\right],\\
		\sum_i\mathbf c_i {H_{1,i}}&=\partial_t(c_1\mathbf{u})+c_s^2 \frac{4\left(1-\phi\right)c_1}{W}\mathbf{n},
		\sum_i \widetilde H_{1,i}=-D_m\nabla\phi\cdot\nabla (c_1+K_{eq}c_2).
	\end{aligned}
\end{equation}
Then we have
\begin{equation}
	\begin{aligned}\label{sum}
		\sum_i h_{1,i}^{(1)}=&-\frac{\delta t}{2} M_{H_{1,i}}^{(1)}, 
		\sum_i h_{1,i}^{(2)}=-\frac{\delta t}{2} M_{H_{1,i}}^{(2)}, \\
		\sum_i H_{1,i}^{(1)}&=M_{H_{1,i}}^{(1)}, 
		\sum_i H_{1,i}^{(2)}=M_{H_{1,i}}^{(2)}.
	\end{aligned}
\end{equation}
Applying Taylor expansion to LB evolution Eq. (\ref{LB_two_1}), one can derive
\begin{equation}
	D_ih_{1,i}+\frac{\delta t}{2}{D_i}^2h_{1,i}+\cdots=-\frac{1}{\tau_{h_{1}} \delta t}(h_{1,i}-h_{1,i}^{eq})+(1-\frac{1}{2\tau_{h_{1}}})H_{1,i}+ \widetilde H_{1,i}(x,t),
\end{equation}
where $D_i=\partial_t+\mathbf c_i  \cdot \nabla$. Substituting Eq. (\ref{expansion}) into above equation, one can get the following equations at different orders of $\epsilon$,
\begin{subequations}\label{20}
	\begin{equation}\label{5a}
		h_{1,i}^{(0)}=h_{1,i}^{eq},
	\end{equation}
	\begin{equation}\label{5b}
		D_{1i}h_{1,i}^{(0)}=-\frac{1}{\tau_{h_{1}} \delta t}h_{1,i}^{(1)}+(1-\frac{1}{2\tau_{h_{1}}})H_{1,i}^{(1)},
	\end{equation}
	\begin{equation}\label{5c}
		\partial_{t_2}h_{1,i}^{(0)}+D_{1i}h_{1,i}^{(1)}+\frac{\delta t}{2}D_{1i}^2 h_{1,i}^{(0)}=-\frac{1}{\tau_{h_{1}} \delta t}h_{1,i}^{(2)}+(1-\frac{1}{2\tau_{h_{1}}})H_{1,i}^{(2)}+{\widetilde H_{1,i}}^{(2)},
	\end{equation}
\end{subequations}
where $D_{1i}=\partial_{t_1}+\mathbf c_i \cdot \nabla_1$. Summing Eq. (\ref{5b}) over $i$ yields
\begin{equation}\label{1epsilon}
	\partial_{t_1}c_1+\nabla_1 \cdot (c_1\mathbf{u})=M_{H_{1,i}}^{(1)}.
\end{equation}
From Eq. (\ref{5b}), we have
\begin{equation}\label{21}
		\frac{\delta t}{2}D_{1i}^2 h_{1,i}^{(0)}=-\frac{1}{2\tau_{h_{1}}}D_{1i}h_{1,i}^{(1)}+\frac{\delta t}{2}(1-\frac{1}{2\tau_{h_{1}}})D_{1i}H_{1,i}^{(1)}.
\end{equation}
Substituting Eq. (\ref{21}) into Eq. (\ref{5c}), one can obtain the following equation,
\begin{equation}\label{19}
	\partial_{t_2}h_{1,i}^{(0)}+(1-\frac{1}{2\tau_{h_{1}}})D_{1i}h_{1,i}^{(1)}+\frac{\delta t}{2}(1-\frac{1}{2\tau_{h_{1}}})D_{1i}H_{1,i}^{(1)}=-\frac{1}{\tau_{h_{1}} \delta t}h_{1,i}^{(2)}+(1-\frac{1}{2\tau_{h_{1}}})H_{1,i}^{(2)}+{\widetilde H_{1,i}}^{(2)}.
\end{equation}
Based on Eq. (\ref{5b}), we also have
\begin{equation}\label{23}
	\mathbf c_i h_{1,i}^{(1)}=\tau_{h_{1}} \delta t \left[ -\mathbf c_i D_{1i} h_{1,i}^{(0)}+ (1-\frac{1}{2\tau_{h_{1}}})\mathbf c_i H_{1,i}^{(1)}\right],
\end{equation}
substituting Eq. (\ref{23}) into Eq. (\ref{19}), one can derive
\begin{equation}
	\begin{aligned}\label{24}
		\partial_{t_2}h_{1,i}^{(0)}=\nabla_1 \cdot \left[(\tau_{h_{1}}-\frac{1}{2})\delta t \left( \mathbf c_i  D_{1i} h_{1,i}^{(0)}-\mathbf c_i H_{1,i}^{(1)}\right)\right]-\frac{1}{\tau_{h_{1}} \delta t}h_{1,i}^{(2)}+(1-\frac{1}{2\tau_{h_{1}}})H_{1,i}^{(2)}+{\widetilde H_{1,i}}^{(2)}.
	\end{aligned}
\end{equation}
Summing Eq. (\ref{24}) over $i$ and applying the conditions in Eqs. (\ref{expansion}) and (\ref{sum}), we get
\begin{equation}
	\begin{aligned}\label{2epsilon}
		\partial_{t_2}c_1=\nabla_1 \cdot \left[D_1 \left( \nabla_1 c_1 -\frac{4\left(1-\phi\right)c_1}{W}{\frac{\nabla_1 \phi}{|\nabla \phi|}} \right)\right]+M_{H_{1,i}}^{(2)}-D_m\nabla_1\phi\cdot\nabla_1 (c_1+K_{eq}c_2),
	\end{aligned}
\end{equation}
where $D_1=\delta t(\tau_{h_{1}}-\frac{1}{2})c_s^2$. Finally, combining Eqs. (\ref{1epsilon}) and (\ref{2epsilon}), one can recover the following governing equation,
\begin{equation}
	\partial_{t}c_1+\nabla \cdot (c_1\mathbf{u})=\nabla \cdot \left[D_1 \left( \nabla c_1 -\frac{4\left(1-\phi\right)c_1}{W}{\mathbf{n}} \right)\right]+AD_m \left[K_{eq}c_2\phi-c_1(1-\phi)\right]-D_m\nabla\phi\cdot\nabla (c_1+K_{eq}c_2).
\end{equation}

\bibliography{reference}
\end{document}